\documentclass[10pt,conference]{IEEEtran}
\IEEEoverridecommandlockouts
\usepackage{cite}
\usepackage{amsmath,amssymb,amsfonts}
\usepackage{algorithmic}
\usepackage{graphicx}
\usepackage{textcomp}
\usepackage{xcolor}
\usepackage{amsthm}
\usepackage{subfigure}
\usepackage{colortbl}
\usepackage{soul}  
\usepackage{enumitem}
\usepackage{listings}
\usepackage{balance}
\usepackage{xcolor}
\definecolor{skyblue}{rgb}{0., 0.72, 0.92}
\usepackage{amsmath, amsfonts}
\usepackage[linesnumbered,ruled,vlined]{algorithm2e}
\SetKwInOut{Input}{Input}
\SetKwInOut{Output}{Output}

\SetCommentSty{mycommfont}

\usepackage{multirow}
\usepackage{booktabs}
\usepackage{fancyhdr}

\theoremstyle{definition}
\newtheorem{definition}{Definition}[]

\definecolor{lightgray}{rgb}{0.78, 0.78, 0.78}
\sethlcolor{lightgray}  
\definecolor{answercolor}{RGB}{240, 240, 240}

\def\BibTeX{{\rm B\kern-.05em{\sc i\kern-.025em b}\kern-.08em
    T\kern-.1667em\lower.7ex\hbox{E}\kern-.125emX}}
\begin{document}


\title{\textit{Aries}: Efficient Testing of Deep Neural Networks via Labeling-Free Accuracy Estimation}

\author{\IEEEauthorblockN{Qiang Hu\IEEEauthorrefmark{2},
Yuejun Guo\thanks{*Corresponding author.}\IEEEauthorrefmark{3}\IEEEauthorrefmark{1}, 
Xiaofei Xie\IEEEauthorrefmark{4}, 
Maxime Cordy\IEEEauthorrefmark{2},
Mike Papadakis\IEEEauthorrefmark{2},
Lei Ma\IEEEauthorrefmark{5}\IEEEauthorrefmark{6} and
Yves Le Traon\IEEEauthorrefmark{2}}
\IEEEauthorblockA{\IEEEauthorrefmark{2}University of Luxembourg, Luxembourg\\
\IEEEauthorrefmark{3}Luxembourg Institute of Science and Technology, Luxembourg \\
\IEEEauthorrefmark{4}Singapore Management University, Singapore  \\
\IEEEauthorrefmark{5}University of Alberta, Canada \\
\IEEEauthorrefmark{6}The University of Tokyo, Japan
}}

\maketitle
\thispagestyle{fancy}
\pagestyle{fancy}
\cfoot{\thepage}
\renewcommand{\headrulewidth}{0pt} 
\renewcommand{\footrulewidth}{0pt}

\begin{abstract}
Deep learning (DL) plays a more and more important role in our daily life due to its competitive performance in industrial application domains. As the core of DL-enabled systems, deep neural networks (DNNs) need to be carefully evaluated to ensure the produced models match the expected requirements. In practice, the \emph{de facto standard} to assess the quality of DNNs in the industry is to check their performance (accuracy) on a collected set of labeled test data. However, preparing such labeled data is often not easy partly because of the huge labeling effort, i.e., data labeling is labor-intensive, especially with the massive new incoming unlabeled data every day. Recent studies show that test selection for DNN is a promising direction that tackles this issue by selecting minimal representative data to label and using these data to assess the model. However, it still requires human effort and cannot be automatic. 
In this paper, we propose a novel technique, named \textit{Aries}, that can estimate the performance of DNNs on new unlabeled data using only the information obtained from the original test data. The key insight behind our technique is that the model should have similar prediction accuracy on the data which have similar distances to the decision boundary. We performed a large-scale evaluation of our technique on two famous datasets, CIFAR-10 and Tiny-ImageNet, four widely studied DNN models including ResNet101 and DenseNet121, and 13 types of data transformation methods. Results show that the estimated accuracy by \textit{Aries} is only 0.03\% -- 2.60\% off the true accuracy. Besides, \textit{Aries} also outperforms the state-of-the-art labeling-free methods in 50 out of 52 cases and selection-labeling-based methods in 96 out of 128 cases.
\end{abstract}


\section{Introduction}
Deep learning (DL) has been continuously deployed and applied in different industrial domains that impact our social society and daily life, such as face recognition~\cite{sun2015deepid3, wang2021deep}, autonomous driving~\cite{grigorescu2020survey, muhammad2020deep}, and video gaming~\cite{vinyals2019grandmaster, ye2020towards}. As the core of DL-enabled systems, deep neural network (DNN) follows the data-driven development paradigm and learns the decision logic automatically based on the incorporated learned knowledge of training data. Similar to traditional software that needs to be well tested, DNNs also need to be comprehensively evaluated before deployment to reduce potential risks in the real world~\cite{tian2018deeptest, li2021testing}.

A common \emph{de facto} standard to assess the quality of DNNs in the industry is by evaluating DNNs on a collected set of labeled data. In practice, when building a DNN model, developers often split a dataset into the training set, validation set, and test set. The test set is mainly used to measure the accuracy of the trained model (as an indicator of performance generality), thus, the final developed DNN often comes with the reported test accuracy. However, the original test set often only covers a part of the data distribution (generally, the same distribution as the training data). The distribution of unseen data is often unclear in the practical context, and the reported test accuracy is hard to reflect the actual model performance in real usage. Therefore, in addition to testing models on the original test data, it is highly desirable to conduct performance evaluation of DNNs on new data, which is generally available from a large amount of daily or monthly incoming data.

However, different from the original test data that already have been labeled, the unseen/new data are usually raw with the absence of label information. It is challenging to assess a model on unlabeled data. More importantly, labeling all the new data (that could be large in size) is labor-intensive and time-consuming, which is almost impossible and impractical. For some complex tasks, domain knowledge from experts is mandatory, leading to the labeling even harder. For example, it takes more than 600 man-hours for experienced software developers to label all the codes from 4 libraries~\cite{NEURIPS2019_49265d24}. 

Towards addressing the data labeling issue for more efficient DNN testing, recently, researchers have adapted the test selection concept~\cite{rothermel1997safe, engstrom2010systematic} from the software engineering community to select and label a subset of representative data, then test and assess the model accordingly. For example, Li \emph{et al.} proposed cross entropy-based sampling (CES)~\cite{li2019boosting} to select a subset that has the minimum cross-entropy with the entire test dataset to test the DNN. In this way, the labeling effort can be significantly reduced and the testing has acceptable bias. However, although test selection is a promising direction for efficient DNN testing, the labeling efforts and costs persist. To bridge the gap, in this paper, we aim to automatically estimate the test accuracy without extra manual labeling.

To this end, we propose a novel technique, \textit{Aries}, to efficiently estimate the performance of DNNs on the new unseen data based on existing labeled test data. The intuition behind our technique is:
\textbf{there can be a connection between the prediction accuracy of the data and the distance of the data from the decision boundary.} More specifically, given two datasets whose distribution of the distance to the decision boundary is close, they could share a similar prediction accuracy. A preliminary empirical study is first conducted to validate our assumption. Specifically, we adopt the existing dropout-based uncertainty to estimate the distance between the data instances and the decision boundary. By splitting the uncertainty score into $n$ intervals, we obtain $n$ buckets of the distance distribution. Then, we can map the data instances into a bucket based on their uncertainty scores. With this, we can estimate the accuracy of each bucket based on the original test data (with labels) that fall into this bucket as the supporting evidence (points). Finally, given the new data without labels, we map them into different buckets and leverage the estimated bucket accuracy to calculate the overall accuracy of the new data. Compared to existing techniques that need the labeled data to calculate the model accuracy, \textit{Aries} is fully automatic without extra human labeling effort.

To assess the effectiveness of \textit{Aries}, we conduct in-depth evaluations on two commonly used datasets, CIFAR-10 and Tiny-ImageNet, four different DNN architectures, including ResNet101 and DenseNet121. Besides the original test data, we also use 13 types of transformed test sets (e.g., data with added brightness) to simulate the new unlabeled data that could occur in the wild, where the transformed data could follow different data distributions from the original test data~\cite{hu2022empirical}. The results demonstrated that \textit{Aries} could precisely estimate the performance of DNN models on new unlabeled data without further labeling effort. Compared to the real accuracy, the estimated accuracy exhibits a difference ranging from 0.03\% to 2.60\% by using the default parameter setting. Besides, compared to the state-of-the-art (SOTA) labeling-free model performance estimation methods~\cite{deng2021labels}, \textit{Aries} can predict more accurate accuracy. And compared to the test selection methods CES~\cite{li2019boosting} and PACE~\cite{chen2020practical}, which need to label a portion of test data, \textit{Aries} can still achieve competitive results without extra labeling. Moreover, we conduct ablation studies to explore the impact of each component of \textit{Aries} on the estimation performance. 

To summarize, the main contributions of this paper are:

\begin{itemize}[leftmargin=*]
    \item We propose a novel DNN testing technique for quality assessment, \textit{Aries}, that can efficiently estimate the model accuracy on unlabeled data without any labeling effort. 
    \item We empirically demonstrate that \textit{Aries} can achieve better results than SOTA labeling-free model performance estimation methods and competitive results compared to test selection methods that require human labeling effort.
    \item We also comprehensively explore each potential factor that could affect the performance of \textit{Aries} and provide the recommendation parameters. 
    \item We release all our source code publicly available\footnote{https://github.com/wellido/Aries\label{site}}, hoping to facilitate further research in this direction.
\end{itemize}

\section{Background}
\label{sec:background}
\subsection{DNN Testing}
DNN testing~\cite{zhang2020machine} is an essential activity in the DL-enabled software development process to ensure the quality and reliability of DNN before deployment. Generally, a deep neural network (DNN) is trained using a large number of labeled data, the so-called training set, with a validation set to estimate the performance (accuracy in this paper) error. Usually, the term ``validation set'' is used interchangeably with ``test set'' given the assumption that the validation set and the test set are derived from the same data distribution as the training set. A minor difference is that the validation set is mostly used in the training process to search for better training settings and configurations. Test data plays the role of future unseen data to estimate how the trained DNN performs in future unseen cases. Although the fundamental assumption of modern machine learning is that, the test (unseen) data and training data share a similar distribution, under which the performance obtained on training data could also generalize to the test data, such an assumption often does not hold for DNNs deployed in the wilds. 
For many real-world applications, the new test set (i.e., future unseen data) can hold a different distribution~\cite{pmlr-v139-koh21a} that undermines the confidence of the obtained accuracy. For example, given a DNN trained on an image set with low contrast, the newly selected images are with high contrast~\cite{hendrycks2019benchmarking}. As a result, the accuracy of the original test set cannot reflect the actual accuracy of the new test set. Moreover, the new data are usually raw and unlabeled to directly compute the accuracy, which requires the developers' dedicated testing. For simplicity, in this paper, we use  ``original test data'' to represent the labeled test data that are accompanied by the training set and ``new unlabeled data'' to indicate the unlabeled test set.

\subsection{Test Optimization in DNN Testing}
Given an unlabeled test set, the most straightforward way to obtain the DNN accuracy is to manually label each data and compare the difference between the ground truth and predicted labels. However, the labeling effort can be costly in labor expense and time. As mentioned in the literature, test selection~\cite{chen2020practical}, has been demonstrated as a promising solution to optimize the testing process. 
Test selection techniques can be divided into two categories: 1)The first one is test prioritization, which tries to identify the data most likely to be misclassified~\cite{prior2021}. After finding these data, they can be used to further enhance the pre-trained model (e.g., by model retraining). 2) The other one is to select a subset of data that can reflect the behavior of the whole set~\cite{chen2020practical}. In this technique, a fixed number of relevant test data are selected via a specific method and manually labeled to calculate the accuracy. The selected set, by default, is (expected to be) representative of the entire set and, thus, the obtained accuracy can approximately reflect the DNN accuracy on the whole given set. Although in this way, the labeling effort can be greatly reduced to a few data, it can still be impractical are too large to be handled under a given budget (e.g., labeling time, available cost). In this paper, our proposed technique falls into the second category. 

\section{Methodology}
\label{sec:methodology}
We first introduce the insight and assumption of \textit{Aries}, then conduct preliminary studies to empirically validate our assumptions, and finally present the details of \textit{Aries}.

\subsection{Key Insight and Assumption}
\begin{figure}[]
	\centering
	\includegraphics[width=0.4\textwidth]{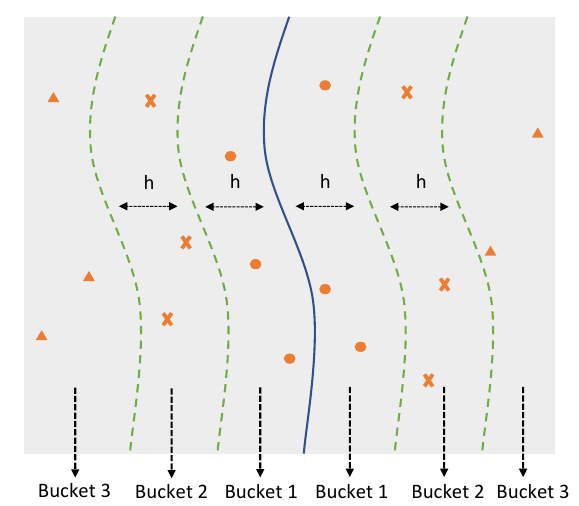}
	\caption{An example of the assumption of our technique. Data in the same bucket are highlighted with the same marker.}
	\label{fig:assumption}
\end{figure}

Our assumption is that there could be a correlation between the prediction accuracy and the distance of the data from the decision boundary. To better understand this assumption, Fig.~\ref{fig:assumption} depicts an intuitive example of a binary classification with a decision boundary (blue solid lines) splitting the data space into 2 regions. Since the data falling into the same region will be predicted by the same label, but with different confidences, we split each region further into multiple sub-regions based on its distance to the boundary. In the figure, we evenly split the space into 3 buckets (\textit{Bucket1}, \textit{Bucket2}, and \textit{Bucket3}) by the distance \textit{h}. We assume that, for data in the same bucket, the probability of making the (in)correct prediction is similar, i.e., the model has the same performance on these data. Thus, if we can obtain the model accuracy in each bucket and map new data into the corresponding bucket, we can approximate the model accuracy on these new data.

The essential insight of our assumption is to measure the distance between data and decision boundaries (i.e., how to define \textit{h} in Fig.~\ref{fig:assumption}). Given the fact that the data space is usually high dimensional and complex, it is difficult to directly describe the decision boundary. Recently, the dropout uncertainty~\cite{gal2016dropout, zhang2020towards} has been widely used to estimate the distance of the data from the decision boundary. Roughly speaking, given one data and a model with dropout layers. After using this model to predict the data multiple times, if the predicted outputs have a huge variance, we say this data might be uncertain by the model and near the decision boundaries. We employ the dropout uncertainty in \textit{Aries} for the distance approximation and will explore other options in the configuration study. At a high level, our insight is sound and practical in the way that the prediction confidence of a DNN follows a particular distribution that is automatically learned from the training data in the training process. When a test sample falls into a distribution and prediction confidence region, our fine-grained split region could provide a certain level of evidence to support DNN quality assessment.

\subsection{Preliminary Study}
\label{subsec:pre}
First, we define $Label \  Variation \  Ratio (LVR)$ to approximate the distance between data and decision boundaries. 

\begin{definition}[\textbf{Label Variation Ratio (LVR)}]
\label{def: lvr}
Given a model $M$ with dropout layers and an input data $x$, the number of dropout predictions $T$, LVR of $x$ is defined as:

$$LVR\left(M,x,T\right)=\frac{\left|\left\{k\mid { 1 \leq  k\leq T\wedge L_{M^k\left(x\right)}={L_{max}}}\right\}\right|}{T}$$

\noindent where $L_{M^k(x)}$ is the $k$-th predicted label of $x$ by $M$ and $L_{max}$ is the dominant label of $T$ predictions (i.e., the label predicted by most predictions). 
Intuitively, the prediction of the data near the boundary has low confidence, i.e., with lower $LVR$.

We then divide the data space into $n$ buckets by splitting the $LVR$ into $n$ equal intervals, where the range of $LVR$ is (0,1]. For example, if $n$ is 2, then we have two buckets, and the corresponding $LVR$ intervals are (0, 0.5] and (0.5, 1]. A data instance falls into a region based on its LVR value. 
\end{definition}

\begin{definition}[\textbf{Bucket}] 
\label{def: area}
Given a dataset $X$, $M$ and the number of intervals $n$, we define the data that belong to $t^{th}$ bucket as:

$$Bucket(X, t, T)=\left\{x \mid x\in X \wedge \frac{t}{n}< LVR(M,x, T) \le \frac{t+1}{n}\right\}$$
where $0\le t<n$ and $T$ is the number of dropout predictions.
\end{definition}

\begin{figure}[!h]
    \centering
    \subfigure[CIFAR10-ResNet20]{
    \includegraphics[scale=0.25]{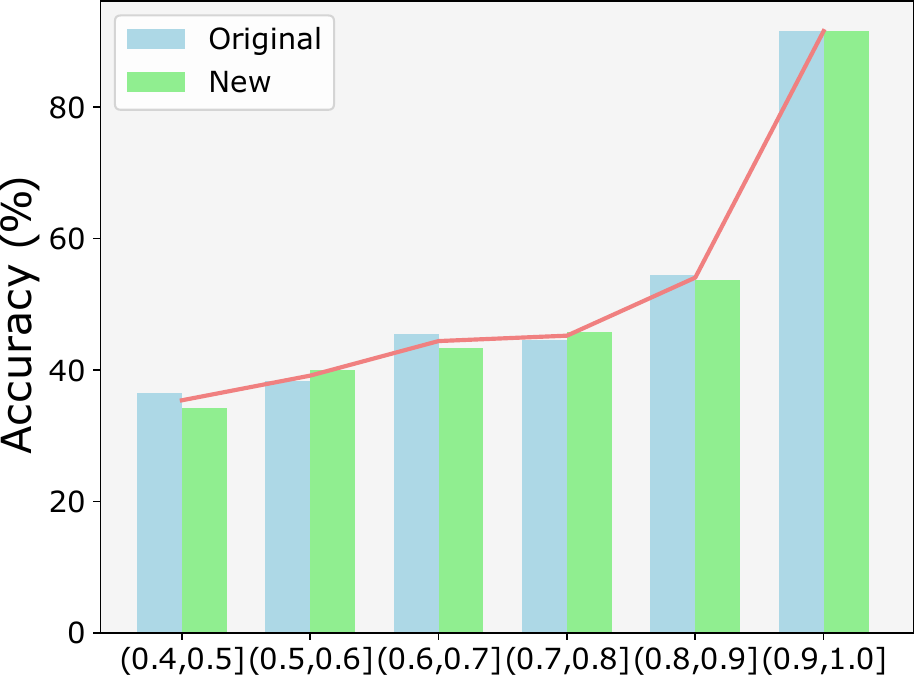}%
    }
    \subfigure[CIFAR10-VGG16]{
    \includegraphics[scale=0.25]{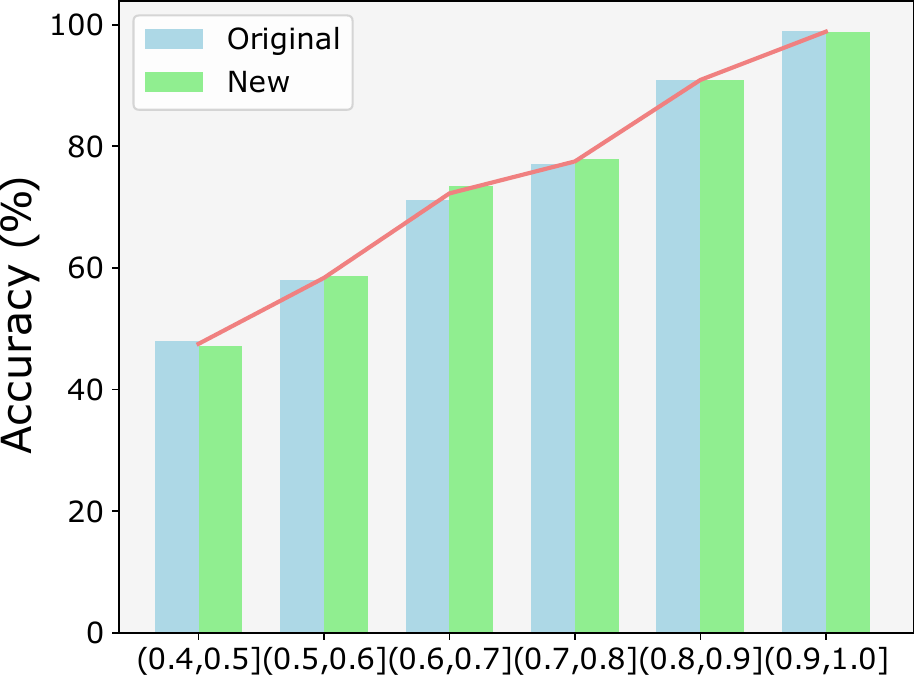}%
    }
    \subfigure[ImageNet-ResNet101]{
    \includegraphics[scale=0.25]{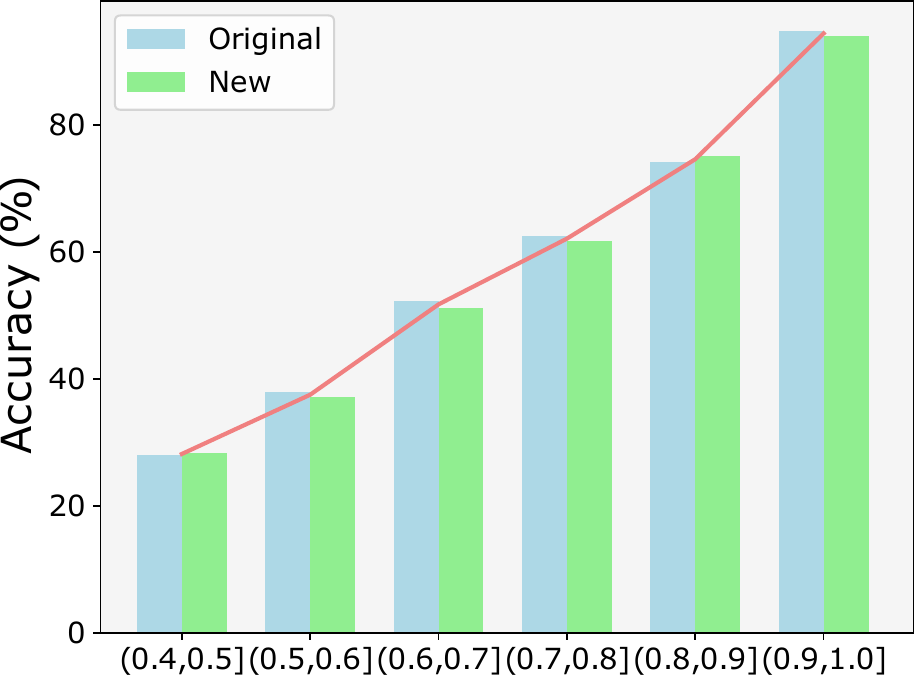}%
    }
    \subfigure[ImageNet-DenseNet121]{
    \includegraphics[scale=0.25]{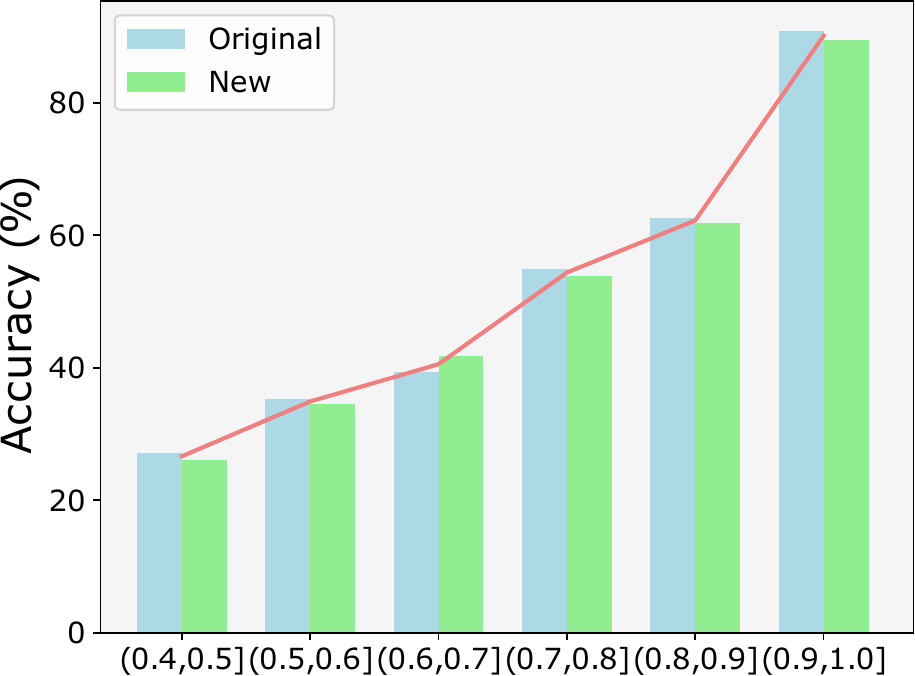}%
    }
    \caption{Accuracy in different $Buckets$. Original: labeled test data, New: unlabeled new unlabeled data.}
    \label{fig:pre_acc_each_boundary}
\end{figure}

Next, to validate the rationality of our assumption, we conduct two preliminary studies to check 1) if data in the same $Bucket$ have similar accuracy, and 2) if there is a relation between $LVR$ and model accuracy (please refer to Section~\ref{sec:setup} for details of datasets and models). In the first study. We randomly split the test data into two sets and assume one set we already have the label information and the other is the new unlabeled data. Then we activate the Dropout layers in each model to predict these two sets multiple times (here, we set the number of $T$ as 50, which is the default setting of \textit{Aries}). Afterward, we calculate the $LVR$ score of each data and then split the data into different $Buckets$ using Definition~\ref{def: area}. Finally, we check the accuracy of the model on the data that are in the same $Bucket$. Fig.~\ref{fig:pre_acc_each_boundary} presents the results of this study. Note that we eliminate the results of data whose $LVR$ scores are smaller than 0.4 due to the negligible amount (e.g., only one in CIFAR-10, ResNet20). We can see that 1) the two sets of data have similar accuracy in the $Buckets$ regardless of the datasets and models, 2) data with higher $LVR$ scores (closer to the decision boundary) have higher accuracy. This finding justifies our assumption.

\noindent\colorbox{white!20}{\framebox{\parbox{0.96\linewidth}{
\textbf{Finding 1}: A DNN has similar accuracy on the data sets that have similar distances to the decision boundary.}}}

\begin{figure}[!ht]
    \centering
    \subfigure[CIFAR10-ResNet20]{
    \includegraphics[scale=0.25]{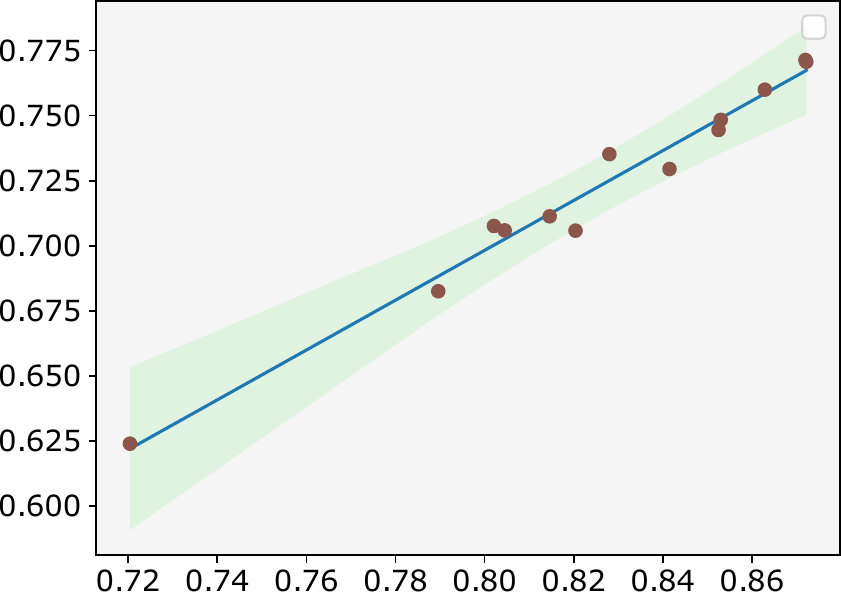}%
    }
    \subfigure[CIFAR10-VGG16]{
    \includegraphics[scale=0.25]{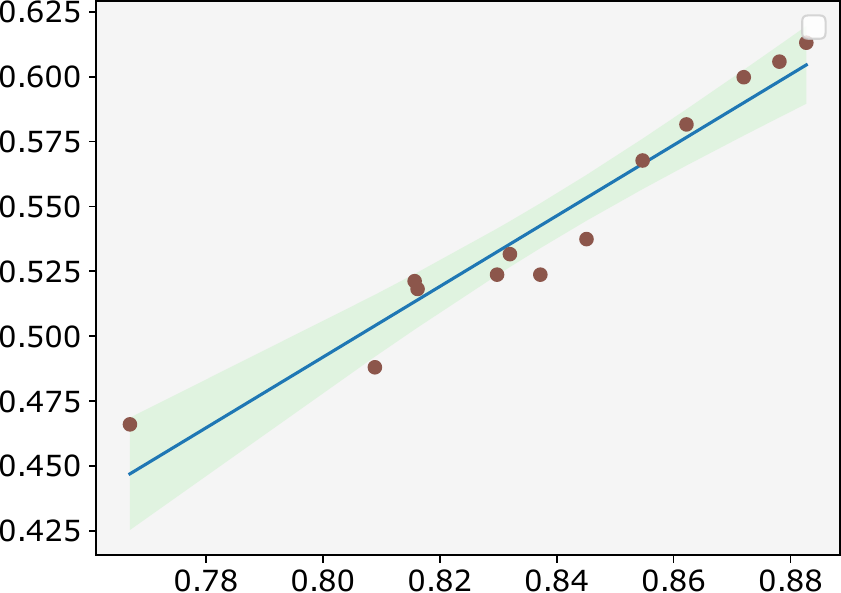}%
    }
    \subfigure[ImageNet-ResNet101]{
    \includegraphics[scale=0.25]{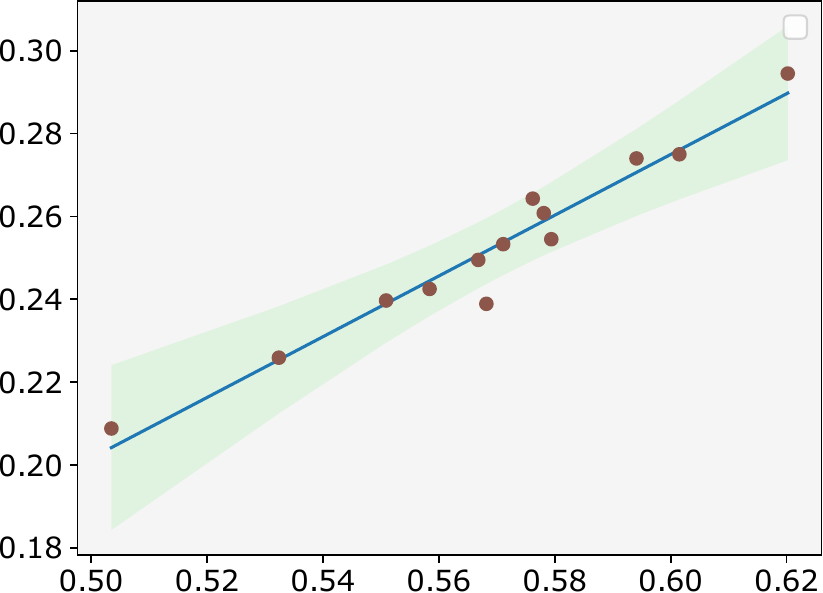}%
    }
    \subfigure[ImageNet-DenseNet121]{
    \includegraphics[scale=0.25]{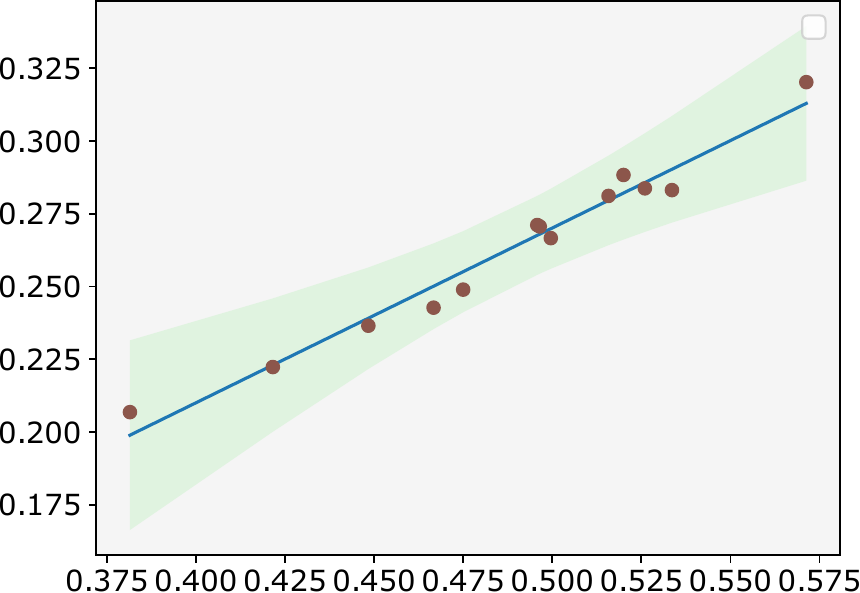}%
    }
    \caption{The linear relation (blue line) between the size of data with the highest label variation ratio ($y$-axis, unit: ratio) and the test accuracy ($x$-axis, unit: \%). We apply the least squares polynomial fit to draw the blue line in each figure.}
    \label{fig:pre_realtion}
\end{figure}

Second, we investigate if there is a relation between the size of data with high $LVR$ and the accuracy of the model on this set. Usually, data with a high $LVR$ score means the model is confident in predicting this data. Intuitively, the size of data that the model has high confidence could partly reflect the model's performance. Fig.~\ref{fig:pre_realtion} depicts the model accuracy (x-axis) and the percentage of highly confident data (y-axis) on each data set. Here, the highly confident data means their $LVR$ is 1. We can see that there is a clear linear relationship between the two values. The results lead to our basic idea:  we can measure the percentage of highly confident data in the new unlabeled data although we do not know the truth labels of the new data. Then, based on existing test data with truth labels, we can measure the accuracy of the datasets with a certain ratio of highly confident data. Finally, we can estimate the accuracy of unlabeled data.

\noindent\colorbox{white!20}{\framebox{\parbox{0.96\linewidth}{
\textbf{Finding 2}: There is a linear relationship between the \% of highly confident data ($LVR=1$) and the accuracy of the whole set.
Therefore, given some labeled data, if we know 1) the accuracy of the DNN in each $Bucket$, and 2) the percentage of highly confident data, it is promising to estimate the accuracy of the new unlabeled data. 
}}}

\subsection{\textit{Aries}: Efficient Testing of DNNs}
\label{sec:pe}
\begin{algorithm}[htpb]
\small
\caption{\textit{Aries}: efficient testing of DNNs}
\label{alg:pe}
\SetAlgoLined
\Input{$M$: model with dropout layer\\
$X_{ori}$: original test data with labels\\
$X_{new}$: new data without labels\\
$T$: number of dropout predictions \\
$n$: number of buckets}
\Output{$Acc_{Aries}$: estimated accuracy of $X_{new}$}
$correct\_num = 0$ \\
\For{$i=0 \to n-1$}
{

    $BucketAcc_i = evaluate(M, Bucket(X_{ori}, i, T))$\;
    $correct\_num += |Bucket(X_{new}, i, T)| \times BucketAcc_i$\;
}
$Acc_{bucket} = correct\_num / len\left(X_{new}\right)$ \;
$Acc_{ori}=evaluate(M, X_{ori})$\;
$Acc_{confident} = \frac{|Bucket(X_{new},n-1)|/len\left(X_{new}\right)}{|Bucket(X_{ori},n-1)|/len\left(X_{ori}\right)} \times Acc_{ori}$ \;
$Acc_{Aries} = average\left(Acc_{bucket}, \  Acc_{confident}\right)$ \;
\Return $Acc_{Aries}$\;
\end{algorithm}

Based on the two findings, we propose a novel technique, \textit{Aries}, that can test the performance of DNN models without requiring the label information. The key idea of \textit{Aries} is to take advantage of the labeled test sets to approximate the model performance on the new unlabeled data. Algorithm~\ref{alg:pe} presents the details of our technique which contains two main steps. 

First, given a model $M$ with dropout layers, the original labeled data $X_{ori}$, the number of buckets $n$, and the dropout prediction time  $T$, \textit{Aries} splits $X_{ori}$ into $n$ buckets according to Definition~\ref{def: area}, and calculate the accuracy of the data in each bucket $BucketAcc_{i}$ (lines 2 and 3). Then, the same as the $X_{ori}$, we split $X_{new}$ into $n$ buckets and use the bucket accuracy $BucketAcc_{i}$ to estimate the correctly predicted number $correct\_num$ of $X_{new}$ (line 4).

Then, we perform the accuracy estimation. First, according to finding 1, \textit{Aries} directly computes the accuracy using the correct data number of new data in each bucket and produces the first estimation $Acc_{bucket}$ (line 6). Then, based on finding 2, we calculate the accuracy of the labeled data first $Acc_{ori}$ (line 7). Afterward, \textit{Aries} computes the proportion of the high confident data in the new unlabeled data to the original test data, and then estimates the second accuracy $Acc_{confident}$ (line 8). In the end, since in practice, the $Acc_{bucket}$ ($Acc_{confident}$) over(under)-estimates the accuracy, we compute the average of the two estimated accuracy as the final output of \textit{Aries}, $Acc_{Aries}$(line 9). We will detail explain why we combine $Acc_{bucket}$ and $Acc_{confident}$ in section~\ref{sec:results}. Highlight that, \textit{Aries} only requires N forward propagations to work, which is significantly more efficient than the existing training-based methods~\cite{deng2021labels}.

\textbf{Example:} Taking the cases in Fig.~\ref{fig:assumption} as an example with 3 Buckets ($n=3$), $Bucket1$, $Bucket2$, and $Bucket3$. We assume that the accuracy of original test data ($Acc_{map}$) in $Bucket1$, $Bucket2$, and $Bucket3$ are 60\%, 70\%, and 80\%, respectively. And the number of original test data ($BucketSize_{ori}$) in each part is 200, 300, and 400, respectively. Then, for the new unlabeled data, the number of data in each section ($BucketSize_{new}$) is 300, 400, and 500. Then the $Acc_1$ is calculated by $(200 \times 60\% + 300 \times 70\% + 400 \times 80\%) / (200 + 300 + 400) = 72.22\%$. Next, assume that the accuracy of the original test data is $70\%$, the $Acc_2$ is calculated by $((500 / 1200) / (400 / 900)) \times 70\% = 66.35\%$. The final output of \textit{Aries} is $(72.22\% + 66.35\%) / 2 = 69.29\%$. 

In \textit{Aries}, the dropout prediction plays an important role in estimating the distance to decision boundaries. Therefore, the dropout rate is the first potential influencing factor. Next, the number of buckets that determines the splitting granularity could be the second influencing factor. In addition, since our technique uses dropout prediction and label change times to approximate the distance between the data and the decision boundaries, the distance approximation method is the third influencing factor. In Section~\ref{sec:rq2} we will explore the influence and the best settings of \textit{Aries} for these three factors.

\section{Experimental Setup}
\label{sec:setup}
To evaluate the effectiveness of \textit{Aries}, we conduct experiments on two popular datasets, four DNN architectures, and 13 types of data transformations that are utilized to generate new unlabeled data. Our in-depth evaluation answers the following research questions:

\textbf{RQ1: How effective is \textit{Aries} in accuracy estimation?} 

\textbf{RQ2: How does each component affect and contribute to the effectiveness of \textit{Aries}?} 

\begin{table}[!t]
\caption{Details of datasets and DNNs}
\label{tab:data_model}
\resizebox{\columnwidth}{!}{
\begin{tabular}{
>{\columncolor[HTML]{FFFFFF}}l 
>{\columncolor[HTML]{FFFFFF}}r 
>{\columncolor[HTML]{FFFFFF}}r 
>{\columncolor[HTML]{FFFFFF}}r 
>{\columncolor[HTML]{FFFFFF}}r 
>{\columncolor[HTML]{FFFFFF}}r 
>{\columncolor[HTML]{FFFFFF}}r }
\toprule
\textbf{Dataset} & \textbf{Classes} & \textbf{Training} & \textbf{Test} & \textbf{DNN} & \textbf{Parameters} & \textbf{Accuracy (\%)} \\ \midrule
\cellcolor[HTML]{FFFFFF} & \cellcolor[HTML]{FFFFFF} & \cellcolor[HTML]{FFFFFF} & \cellcolor[HTML]{FFFFFF} & ResNet20 & 274442 & 87.44 \\
\multirow{-2}{*}{\cellcolor[HTML]{FFFFFF}\textbf{CIFAR-10}} & \multirow{-2}{*}{\cellcolor[HTML]{FFFFFF}10} & \multirow{-2}{*}{\cellcolor[HTML]{FFFFFF}50k} & \multirow{-2}{*}{\cellcolor[HTML]{FFFFFF}10k} & VGG16 & 2859338 & 91.39 \\ \midrule
\cellcolor[HTML]{FFFFFF} & \cellcolor[HTML]{FFFFFF} & \cellcolor[HTML]{FFFFFF} & \cellcolor[HTML]{FFFFFF} & ResNet101 & 43036360 & 74.09 \\
\multirow{-2}{*}{\cellcolor[HTML]{FFFFFF}\textbf{Tiny-ImageNet}} & \multirow{-2}{*}{\cellcolor[HTML]{FFFFFF}200} & \multirow{-2}{*}{\cellcolor[HTML]{FFFFFF}100k} & \multirow{-2}{*}{\cellcolor[HTML]{FFFFFF}10k} & DenseNet121 & 7242504 & 70.70 \\ \bottomrule
\end{tabular}
}
\end{table}

\textbf{Subject datasets and DNN models.} Table~\ref{tab:data_model} presents the details of datasets and models. CIFAR-10~\cite{krizhevsky2009learning} is a 10-class dataset of color images, e.g., airplanes and birds. For this dataset, we build two models, ResNet20~\cite{7780459} and VGG16~\cite{simonyan2014very}. Tiny-ImageNet~\cite{le2015tiny} is a more complex dataset that contains 200 image categories, e.g., goldfish and monarch. For this dataset, we use ResNet101 and DenseNet121. In our experiments, we take the original test data for the decision boundary analysis to estimate the accuracy of new data. 

For the new unlabeled data preparation, we directly use the popular natural robustness benchmark dataset~\cite{hendrycks2019benchmarking} for our experiments. This benchmark provides two datasets, CIFAR-10-C and Tiny-ImageNet-C, that are generated by adding common corruptions and perturbations into the original test data, e.g., by increasing the brightness of the image. It is widely used for evaluating the model performance on distribution-shifted data (unseen data). Besides, a recent study~\cite{hu2022empirical} also demonstrates that these kinds of corrupted data can be regarded as out-of-distribution data, because the distance between these data and the original test data is farther than the distance between the real out-of-distribution data and the original test data. In our evaluation, we collect 13 common types that CIFAR-10 and Tiny-ImageNet both include, shown in Table~\ref{tab:data_transformation}.

\begin{table}[]
\caption{Details of data transformation methods used for generating new unlabeled data.}
\label{tab:data_transformation}
\resizebox{\columnwidth}{!}{
\begin{tabular}{ll}
\toprule
\textbf{Data Type} & \textbf{Description} \\ \midrule
Brightness & Increase the brightness of the image data \\ 
Contrast & Increase the contrast of the object \\
Defocus Blur (DB) & Add the defocus blur effect to the image\\
Elastic Transform (ET) & Elastic deformation of images \\
Fog & Add fog effect to the image \\
Frost & Add frost effect to the image\\
Gaussian Noise (GN) & Add Gaussian Noise perturbation to the image\\
Jpeg Compression (JC) & Change the image to Jpeg format \\
Motion Blur (MB) & Add the motion blur effect to the image \\
Pixelate & Convert image to pixelate style \\
Shot Noise (SN) &  Add noise by using the Poisson process\\
Snow &  Add snow effect\\
Zoom Blur (ZB) & Zoom the image data \\ \bottomrule
\end{tabular}
}
\end{table}

\textbf{Baseline.} 
We first compare \textit{Aries} with two SOTA labeling-free model performance estimation methods: 
\begin{enumerate}[leftmargin=*]
\item \textbf{Predicted Score-based Method (Predicted Score)} assumes that a test sample is correctly classified when its maximum output probability is greater than a threshold $\tau$. In the experiments, we follow the same setting as~\cite{deng2021labels} to set the $\tau$ as 0.7, 0.8, and 0.9.
\item \textbf{Meta-set}~\cite{deng2021labels} is recently proposed three-step method. First, Meta-set generates multiple diverse test sets by performing different image transformations on the original test set. Then, it computes the Frechet Distance (FD) between the internal outputs of generated and original test sets. Finally, a regression model is trained by using the FD score and the accuracy of test sets. When new data come, Meta-set calculates its FD score with the original test set and then utilizes the regression model to estimate its accuracy. In the experiments, we utilize both linear regression and neural network regression to build the model. The sizes of the meta set and sample set are set as 1000 and 10000, respectively, following the original paper. In addition, we considered the number of image transformations as 1, 2, and 3, whereas the original paper only studied 3. In this way, we build a strong baseline. 
\end{enumerate}

Then, we compare \textit{Aries} with existing test selection-based model evaluation methods, Cross Entropy-based Sampling (CES)~\cite{li2019boosting} and Practical Accuracy Estimation (PACE)~\cite{chen2020practical}, as baselines. Besides, we also consider random selection as the third baseline. Remarkably, all these three baselines require selecting and labeling a subset from the new unlabeled data to perform testing. We follow the same configuration as the paper~\cite{chen2020practical} to set the labeling budget from 50 to 180 in intervals of 10 for test selection methods. 

\begin{enumerate}[leftmargin=*]
\item \textbf{Cross Entropy-based Sampling (CES)} selects the data that have the minimum cross-entropy with the entire test dataset. It starts from a randomly selected small size of data and iteratively increases the size by adding other data while controlling the cross-entropy.
\item \textbf{Practical Accuracy Estimation (PACE)} first clusters data based on the output of the last hidden layer into different groups by using the hierarchical density-based spatial clustering of applications with noise clustering method, then utilizes the example-based explanation algorithm MMD-critic to select the most representative data from each group to label and test the model.
\end{enumerate}

\textbf{\textit{Aries} configuration.} To reduce the number of hyperparameters in \textit{Aries}, we set the dropout prediction number to be the same as the number of $Buckets$ ($T=n$ in algorithm~\ref{alg:pe}), which means in each $Bucket$, all the data have the same $LVR$ score. Then, there are two remaining hyperparameters we need to set 1) the number of $Buckets$ we split and 2) the dropout rate. The default setting of the number of $Buckets$ and the dropout rate in RQ1 is 50 and 0.5, respectively. In RQ2, we study the different $Bucket$ number settings 10, 50, 100, and 150, and different dropout rate settings 0.1, 0.2, 0.3, 0.4, 0.5, 0.6, 0.7, 0.8, and 0.9. Since the setting space is infinite and it's impossible to study all, we recommend the best one among our studied settings and show that by this setting we can already get acceptable estimation results. 

\begin{table}[]
\caption{Details of our used model-level mutation operators.}
\label{tab:mutation}
\resizebox{\columnwidth}{!}{
\begin{tabular}{lll}
\toprule
\textbf{Level} & \textbf{Operator} & \textbf{Description} \\ \midrule
\multirow{2}{*}{Weight} & Gaussian fuzzing & Fuzz the weights using Gaussian noise \\
 & Weight Shuffle & Shuffle the weights in the same neuron \\ \midrule
\multirow{3}{*}{Neuron} & Neuron Effect Block & Block a neuron effect, i.e., change the output to 0 \\
 & Neuron Activation Inverse & Invert the activation status \\
 & Neuron Switch & Switch two neurons in the same layer \\ \bottomrule
\end{tabular}
}
\end{table}

\textbf{Model mutation.} In RQ2, we investigate if the DNN model mutation~\cite{hu2019deepmutation++, humbatova2021deepcrime} can be used to replace the dropout prediction for the distance of data to boundary approximation. Similar to the dropout, mutation can produce a variant of the original model with a similar accuracy without retraining the model from scratch~\cite{hu2019deepmutation++}. The difference is that the dropout technique tends to fully ignore some randomly selected neurons, while the mutation technique can also work on the weight of neurons and the neuron status. First, we randomly use the weight-level and neuron-level mutation operators provided by~\cite{hu2019deepmutation++} to generate mutants. The detailed information of the operators is presented in table~\ref{tab:mutation}. To preserve the quality of the mutants, we set the mutation ratio as 0.1 (0.01) for CIFAR-10 (Tiny-ImageNet) models, and the accuracy threshold as 0.9. Then, we follow the steps in Algorithm~\ref{alg:pe} to estimate the accuracy of the model on the new unlabeled data.

\textbf{Implementation and environments.} We implement \textit{Aries} in Python based on TensorFlow~\cite{abadi2016tensorflow} framework.
For the baselines CES and PACE, we utilize their original implementation. For model mutation, we use the available mutation framework provided by the authors. We conduct all the experiments on a 2.6 GHz Intel Xeon Gold 6132 CPU with an NVIDIA Tesla V100 16G SXM2 GPU. 
To counteract randomized factors, we repeat all the experiments 5 times and report the average results in this paper. 
Due to the page limit, we put more detailed results as well as the source code for reproducible study of this paper at the companion site.

\section{Results Analysis}
\label{sec:results}
\subsection{RQ1: Effectiveness of \textit{Aries}}
\textbf{Effectiveness on in-distribution data.} First, we evaluate the accuracy estimation effectiveness of the data (the two sets randomly split from the original test data) used in our preliminary study (Section~\ref{subsec:pre}). Table~\ref{pre:acc_es} lists the results. Recall that $Acc_{bucket}$ is the accuracy estimated by the $Bucket$ accuracy. $Acc_{confident}$ is the accuracy estimated by using the size of high confident data. $Acc_{Aries}$ is the final estimated accuracy of $Aries$, the weighted sum of $Acc_{bucket}$ and $Acc_{confident}$. We report $Acc_{bucket}$ and $Acc_{confident}$ to verify the importance of each component of our technique. The results demonstrate that all three estimations can predict the model accuracy on the new data that follow the same data distribution as the original test data with the difference slightly ranging from 0.07\% to 1.02\%. In addition, it's hard to determine which estimation is the best since one can perform better or worse than the others in different datasets and models.

\begin{table}[h]
\caption{Estimated accuracy and the absolute difference between estimated accuracy and real accuracy(\%) on the test data (New) used in the preliminary study. $Real$: real accuracy. $Acc_{bucket}$, $Acc_{confident}$, and $Acc_{Aries}$ refer to lines 6, 8, and 9 in Algorithm~\ref{alg:pe}, respectively. The best is \hl{highlighted}.}
\label{pre:acc_es}
\resizebox{.98\columnwidth}{!}
{
\begin{tabular}{lcc|cc}
\toprule
 & \multicolumn{2}{c|}{\textbf{CIFAR-10}} & \multicolumn{2}{c}{\textbf{Tiny-ImageNet}} \\
 & \multicolumn{1}{l}{\textbf{ResNet20}} & \multicolumn{1}{l|}{\textbf{VGG16}} & \multicolumn{1}{l}{\textbf{ResNet101}} & \multicolumn{1}{l}{\textbf{DenseNet121}} \\ \hline
\textbf{Real} & 87.28 & 91.18 & 74.16 & 71.98 \\
\textbf{$Acc_{bucket}$} & 87.41 (0.13) & 91.33 (0.15) & \cellcolor[HTML]{C0C0C0}73.73 (0.43) & \cellcolor[HTML]{C0C0C0}71.93 (0.05) \\
\textbf{$Acc_{confident}$} & \cellcolor[HTML]{C0C0C0}87.35 (0.07) & 90.81 (0.37) & 73.39 (0.77) & 70.96 (1.02) \\
\textbf{$Acc_{Aries}$} & 87.38 (0.10) & \cellcolor[HTML]{C0C0C0}91.07 (0.11) & 73.56 (0.60) & 71.45 (0.54) \\ \hline
\end{tabular}
}
\end{table}

\begin{table*}[h]
\centering
\caption{Difference between estimated and real accuracy (\%) under data distribution shifts. $Real$: real accuracy,$Acc_{bucket}$, $Acc_{confident}$, and $Acc_{Aries}$ refer to lines 6, 8, and 9 in Algorithm~\ref{alg:pe}, respectively. Meta-set-Linear(NN)-X means Meta-set with linear (neural network) regression and X types of image transformation combination. The best estimation is \hl{highlighted}.} 
\label{table:rq1_robustness}
\resizebox{\textwidth}{!}
{
\begin{tabular}{lllcccccccccccccc}
\toprule
\textbf{Dataset} & \textbf{DNN} &  & \textbf{Brightness} & \textbf{Contrast} & \textbf{DB} & \textbf{ET} & \textbf{Fog} & \textbf{Frost} & \textbf{GN} & \textbf{JC} & \textbf{MB} & \textbf{Pixelate} & \textbf{SN} & \textbf{Snow} & \textbf{ZB} & \multicolumn{1}{l}{\textbf{Avg.}} \\ \hline
 &  & Real & 87.07 & 85.25 & 87.22 & 79.11 & 86.29 & 82.04 & 87.2 & 81.46 & 78.21 & 82.8 & 78.96 & 80.45 & 76.4 & 82.50 \\
 &  & Predicted score ($\tau$ = 0.7) & 6.48 & 6.72 & 5.81 & 9.78 & 6.22 & 8.59 & 5.82 & 5.58 & 10.76 & 8.92 & 10.51 & 10.49 & 10.09 & 8.14 \\
 &  & Predicted score ($\tau$ = 0.8) & 3.46 & 3.11 & 2.61 & 4.87 & 2.96 & 4.35 & 2.65 & 0.43 & 5.56 & 5.18 & 5.75 & 6.03 & 4.74 & 3.98 \\
 &  & Predicted score ($\tau$ = 0.9) & 1.33 & 2.2 & 1.95 & 2.38 & 1.79 & 1.48 & 1.99 & 8.27 & 1.98 & 0.18 & 0.97 & 0.15 & 3.37 & 2.16 \\
 &  & Meta-set-Linear-1 & 8.91 & 7.86 & 9.16 & 2.76 & 8.41 & 5.38 & 9.14 & 4.56 & 3.48 & 5.65 & 3.75 & 3.69 & 3.00 & 5.83 \\
 &  & Meta-set-Linear-2 & 25.87 & 24.57 & 26.09 & 19.13 & 25.28 & 21.85 & 26.07 & 21.11 & 19.33 & 22.28 & 19.75 & 20.20 & 18.41 & 22.30 \\
 &  & Meta-set-Linear-3 & 25.69 & 24.41 & 25.91 & 19.01 & 25.10 & 21.72 & 25.89 & 20.97 & 19.26 & 22.13 & 19.67 & 20.06 & 18.39 & 22.17 \\
 &  & Meta-set-NN-1 & 5.11 & 7.99 & 4.51 & 4.85 & 8.16 & 10.87 & 4.53 & 9.21 & 3.30 & 8.27 & 11.36 & 6.52 & 2.87 & 6.73 \\
 &  & Meta-set-NN-2 & 7.85 & 11.76 & 8.24 & 11.24 & 11.61 & 12.77 & 8.23 & 10.56 & 11.11 & 6.23 & 10.44 & 8.94 & 11.50 & 10.04 \\
 &  & Meta-set-NN-3 & 3.71 & 2.85 & 3.88 & 5.53 & 1.41 & 7.44 & 3.91 & 3.96 & 4.90 & 3.55 & 5.34 & 4.97 & 7.86 & 4.56 \\
 &  & $Acc_{bucket}$ & 0.26 & 0.90 & \cellcolor[HTML]{C0C0C0}0.09 & 4.03 & 0.42 & 2.73 & \cellcolor[HTML]{C0C0C0}0.10 & 2.31 & 4.33 & 2.91 & 3.75 & 3.24 & 5.04 & 2.32 \\
 &  & $Acc_{confident}$ & \cellcolor[HTML]{C0C0C0}0.23 & 1.25 & 0.59 & 2.58 & 0.83 & 1.89 & 0.60 & 1.20 & 3.86 & \cellcolor[HTML]{C0C0C0}0.16 & 2.27 & 1.36 & 4.97 & 1.68 \\
 & \multirow{-13}{*}{\textbf{ResNet20}} & $Acc_{Aries}$ & 0.25 & \cellcolor[HTML]{C0C0C0}0.18 & 0.34 & \cellcolor[HTML]{C0C0C0}0.72 & \cellcolor[HTML]{C0C0C0}0.21 & \cellcolor[HTML]{C0C0C0}0.42 & 0.35 & \cellcolor[HTML]{C0C0C0}0.55 & \cellcolor[HTML]{C0C0C0}0.23 & \cellcolor[HTML]{FFFFFF}1.54 & \cellcolor[HTML]{C0C0C0}0.74 & \cellcolor[HTML]{C0C0C0}0.94 & \cellcolor[HTML]{C0C0C0}0.04 & \cellcolor[HTML]{C0C0C0}0.50 \\ \cline{2-17} 
 &  & Real & 91.41 & 90.96 & 91.78 & 88.69 & 91.15 & 87.39 & 91.78 & 87.12 & 89.55 & 88.81 & 85.97 & 85.77 & 88.55 & 89.15 \\
 &  & Predicted score ($\tau$ = 0.7) & 4.93 & 4.96 & 4.58 & 6.04 & 4.83 & 7.39 & 4.54 & 7.73 & 6.12 & 6.24 & 7.81 & 8.13 & 6.15 & 6.11 \\
 &  & Predicted score ($\tau$ = 0.8) & 2.65 & 2.57 & 2.62 & 3.41 & 2.59 & 4.61 & 2.58 & 5.24 & 3.82 & 3.85 & 5.33 & 4.89 & 3.56 & 3.67 \\
 &  & Predicted score ($\tau$ = 0.9) & 0.78 & 1.75 & 0.90 & 0.35 & 0.69 & 2.75 & 0.85 & 1.05 & 1.44 & \cellcolor[HTML]{C0C0C0}0.42 & 1.11 & 1.95 & 1.34 & 1.18 \\
 &  & Meta-set-Linear-1 & 6.16 & 6.68 & 6.59 & 5.46 & 6.23 & 3.75 & 6.58 & 2.51 & 7.25 & 3.99 & 2.10 & \cellcolor[HTML]{C0C0C0}1.31 & 8.30 & 5.15 \\
 &  & Meta-set-Linear-2 & 11.66 & 12.10 & 12.08 & 10.79 & 11.70 & 9.11 & 12.07 & 7.95 & 12.50 & 9.45 & 7.48 & 6.74 & 13.37 & 10.54 \\
 &  & Meta-set-Linear-3 & 17.41 & 17.74 & 17.83 & 16.30 & 17.41 & 14.67 & 17.82 & 13.63 & 17.90 & 15.16 & 13.07 & 12.40 & 18.52 & 16.14 \\
 &  & Meta-set-NN-1 & 162.32 & 150.67 & 157.85 & 135.52 & 153.88 & 148.12 & 157.93 & 141.46 & 137.33 & 154.80 & 145.26 & 146.55 & 119.96 & 147.05 \\
 &  & Meta-set-NN-2 & 18.56 & 9.53 & 16.85 & 6.00 & 12.95 & 9.27 & 16.79 & 10.99 & 5.89 & 15.56 & 8.86 & 13.11 & 6.09 & 11.57 \\
 &  & Meta-set-NN-3 & 16.22 & 27.11 & 15.89 & 27.55 & 22.52 & 29.10 & 15.98 & 24.48 & 29.34 & 21.22 & 29.72 & 26.21 & 33.94 & 24.56 \\
 &  & \textbf{$Acc_{bucket}$} & \cellcolor[HTML]{C0C0C0}0.01 & \cellcolor[HTML]{C0C0C0}0.25 & \cellcolor[HTML]{C0C0C0}0.01 & 0.82 & 0.19 & \cellcolor[HTML]{C0C0C0}1.64 & \cellcolor[HTML]{C0C0C0}0.00 & 1.93 & 0.94 & 1.05 & 2.33 & 2.31 & 0.82 & 0.95 \\
 &  & \textbf{$Acc_{confident}$} & 0.65 & 1.79 & 1.47 & 0.99 & 0.51 & 6.85 & 1.26 & 3.38 & 1.99 & 5.68 & 1.78 & 5.21 & 2.45 & 2.62 \\
\multirow{-26}{*}{\textbf{CIFAR-10}} & \multirow{-13}{*}{\textbf{VGG16}} & \textbf{$Acc_{Aries}$} & 0.33 & 0.77 & 0.74 & \cellcolor[HTML]{C0C0C0}0.08 & \cellcolor[HTML]{C0C0C0}0.16 & 2.60 & 0.63 & \cellcolor[HTML]{C0C0C0}0.73 & \cellcolor[HTML]{C0C0C0}0.53 & 2.19 & \cellcolor[HTML]{C0C0C0}0.27 & 1.45 & \cellcolor[HTML]{C0C0C0}0.81 & \cellcolor[HTML]{C0C0C0}0.87 \\ \hline
 &  & Real & 63.93 & 50.18 & 55.77 & 55.33 & 60.23 & 57.6 & 57.96 & 59.58 & 56.86 & 62.37 & 57.17 & 57.81 & 53.15 & 57.53 \\
 &  & Predicted score ($\tau$ = 0.7) & 53.92 & 47.55 & 50.88 & 49.74 & 52.42 & 50.94 & 50.08 & 55.57 & 50.07 & 54.83 & 51.31 & 47.76 & 45.53 & 50.82 \\
 &  & Predicted score ($\tau$ = 0.8) & 60.77 & 51.87 & 55.05 & 54.09 & 57.68 & 56.46 & 56.27 & 57.80 & 55.42 & 59.58 & 55.63 & 55.59 & 53.37 & 56.12 \\
 &  & Predicted score ($\tau$ = 0.9) & 62.99 & 50.10 & 54.61 & 53.81 & 58.69 & 56.75 & 57.02 & 58.82 & 55.56 & 61.52 & 56.23 & 55.85 & 52.52 & 56.50 \\
 &  & Meta-set-Linear-1 & 2.59 & 7.31 & 2.00 & 3.47 & 0.76 & 3.32 & 1.3 & 1.32 & 1.61 & 0.75 & 1.11 & 2.82 & 3.36 & 2.44 \\
 &  & Meta-set-Linear-2 & 19.38 & 5.55 & 11.15 & 10.73 & 15.68 & 13.04 & 13.37 & 15.02 & 12.25 & 17.83 & 12.56 & 13.25 & 8.5 & 12.95 \\
 &  & Meta-set-Linear-3 & 28.84 & 14.99 & 20.58 & 20.17 & 25.13 & 22.5 & 22.81 & 24.48 & 21.69 & 27.29 & 22 & 22.7 & 17.93 & 22.39 \\
 &  & Meta-set-NN-1 & 4.9 & 11.4 & 2.73 & 4.06 & 0.28 & 2.57 & 3.75 & 3.46 & 1.92 & 3.28 & 3.56 & 1.92 & 5.79 & 3.82 \\
 &  & Meta-set-NN-2 & 14.37 & 3.48 & 7.75 & 6.46 & 12.57 & 13.03 & 14.25 & 13.78 & 8.65 & 13.97 & 13.91 & 11.91 & 4.66 & 10.68 \\
 &  & Meta-set-NN-3 & 25.6 & 16.54 & 18.91 & 18.09 & 24.32 & 25.21 & 26.73 & 24.38 & 19.76 & 25.03 & 26.52 & 23.48 & 16.43 & 22.38 \\
 &  & \textbf{$Acc_{bucket}$} & 4.48 & 10.76 & 7.80 & 7.61 & 5.70 & 7.00 & 7.03 & 6.26 & 7.27 & 5.22 & 7.27 & 6.55 & 9.16 & 7.09 \\
 &  & \textbf{$Acc_{confident}$} & 5.06 & 10.83 & 7.03 & 7.45 & 6.81 & 6.44 & 6.58 & 5.97 & 7.12 & 5.32 & 6.66 & 7.73 & 10.25 & 7.17 \\
 & \multirow{-13}{*}{\textbf{ResNet101}} & \textbf{$Acc_{Aries}$} & \cellcolor[HTML]{C0C0C0}0.29 & \cellcolor[HTML]{C0C0C0}0.03 & \cellcolor[HTML]{C0C0C0}0.38 & \cellcolor[HTML]{C0C0C0}0.08 & \cellcolor[HTML]{C0C0C0}0.55 & \cellcolor[HTML]{C0C0C0}0.28 & \cellcolor[HTML]{C0C0C0}0.23 & \cellcolor[HTML]{C0C0C0}0.15 & \cellcolor[HTML]{C0C0C0}0.08 & \cellcolor[HTML]{C0C0C0}0.05 & \cellcolor[HTML]{C0C0C0}0.31 & \cellcolor[HTML]{C0C0C0}0.59 & \cellcolor[HTML]{C0C0C0}0.54 & \cellcolor[HTML]{C0C0C0}0.27 \\ \cline{2-17} 
 &  & Real & 58.26 & 36.19 & 44.9 & 46.94 & 53.58 & 49.83 & 51.85 & 52.37 & 47.54 & 57.4 & 51.26 & 49.74 & 42.28 & 49.40 \\
 &  & Predicted score ($\tau$ = 0.7) & 5.67 & 7.86 & 12.39 & 10.69 & 7.41 & 9.92 & 6.82 & 8.32 & 11.13 & 5.75 & 7.43 & 10.51 & 10.94 & 8.83 \\
 &  & Predicted score ($\tau$ = 0.8) & 15.00 & 4.98 & 10.42 & 12.74 & 13.68 & 12.53 & 17.20 & 13.42 & 11.92 & 15.06 & 14.78 & 12.78 & 9.79 & 12.64 \\
 &  & Predicted score ($\tau$ = 0.9) & 23.84 & 13.03 & 18.01 & 20.66 & 22.68 & 20.88 & 22.64 & 21.62 & 19.69 & 24.14 & 23.05 & 20.99 & 17.50 & 20.67 \\
 &  & Meta-set-Linear-1 & 1.23 & 10.05 & 3.32 & 4.54 & 2.31 & 3.69 & 1.41 & 2.85 & 3.51 & 0.22 & 1.58 & 3.7 & 3.31 & 3.21 \\
 &  & Meta-set-Linear-2 & 19.71 & 2.54 & 6.2 & 8.3 & 15.01 & 11.22 & 13.24 & 13.79 & 8.89 & 18.86 & 12.64 & 11.13 & 3.54 & 11.16 \\
 &  & Meta-set-Linear-3 & 29.07 & 6.7 & 15.46 & 17.59 & 24.36 & 20.54 & 22.55 & 23.13 & 18.18 & 28.23 & 21.95 & 20.45 & 12.77 & 20.08 \\
 &  & Meta-set-NN-1 & 0.54 & 25 & 13.3 & 10.94 & 3.75 & 7.87 & 3.92 & 2.62 & 10.86 & 0.95 & 5.1 & 5.04 & 16.32 & 8.17 \\
 &  & Meta-set-NN-2 & 18.26 & 0.11 & 5.29 & 7.93 & 16.83 & 13.69 & 15.28 & 15.53 & 8.25 & 19.58 & 14.13 & 14.81 & 2.68 & 11.72 \\
 &  & Meta-set-NN-3 & 29.23 & 14.15 & 15.89 & 18.11 & 27.32 & 25.37 & 25.23 & 24.23 & 18.8 & 29.07 & 24.46 & 24.08 & 13.5 & 22.26 \\
 &  & \textbf{$Acc_{bucket}$} & 5.67 & 7.86 & 12.39 & 10.69 & 7.41 & 9.92 & 6.82 & 8.32 & 11.13 & 5.75 & 7.43 & 10.51 & 10.94 & 8.83 \\
 &  & \textbf{$Acc_{confident}$} & 5.80 & 7.11 & 7.42 & 10.98 & 8.14 & 7.66 & 6.06 & 7.32 & 7.33 & 5.68 & 6.31 & 6.68 & 10.12 & 7.43 \\
\multirow{-26}{*}{\textbf{Tiny-ImageNet}} & \multirow{-13}{*}{\textbf{DenseNet121}} & \textbf{$Acc_{Aries}$} & \cellcolor[HTML]{C0C0C0}0.07 & \cellcolor[HTML]{C0C0C0}0.37 & \cellcolor[HTML]{C0C0C0}2.49 & \cellcolor[HTML]{C0C0C0}0.15 & \cellcolor[HTML]{C0C0C0}0.36 & \cellcolor[HTML]{C0C0C0}1.13 & \cellcolor[HTML]{C0C0C0}0.38 & \cellcolor[HTML]{C0C0C0}0.50 & \cellcolor[HTML]{C0C0C0}1.90 & \cellcolor[HTML]{C0C0C0}0.03 & \cellcolor[HTML]{C0C0C0}0.56 & \cellcolor[HTML]{C0C0C0}1.92 & \cellcolor[HTML]{C0C0C0}0.41 & \cellcolor[HTML]{C0C0C0}0.79 \\ \toprule
\end{tabular}
}
\end{table*}

\textbf{Effectiveness on data with distribution shift.} In real-world applications, software developers are more interested in the data that follow various data distributions since after the model has been deployed in the wild, the distribution of new unlabeled data is uncontrollable. Thus, we evaluate \textit{Aries} using the data that contain different data distributions. 
Table~\ref{table:rq1_robustness} summarizes the results of the accuracy estimation on the 13 types of distribution-shifted test sets. The same to the results of the preliminary study, we report all three estimations here to analyze the contribution of each part. Overall, different from the results on the original test data, the $Acc_{Aries}$ is closer to the real accuracy in most cases (42 out of 52) than $Acc_{bucket}$ and $Acc_{confident}$. On average, compared to the real accuracy, the difference of estimated accuracy $Acc_{Aries}$ ranges from 0.03\% to 2.60\% across all the datasets and models. And for each models, the average difference of $Acc_{Aries}$ is smaller than 1\% (0.52\%, 0.91\%, 0.27\%, and 0.85\% for ResNet20, VGG16, ResNet101, and DenseNet121). Surprisingly, for Tiny-ImageNet, ResNet101, which has the most complex model architecture among all the models we considered, all the estimated biases are smaller than 0.59\%. This means, our technique is flexible and still effective on challenging datasets and models. 

More specifically, the results reveal that, when we utilize \textit{Aries} to estimate the accuracy of distribution shifted data, only considering the $Acc_{bucket}$ or $Acc_{confident}$ is not enough, especially on the complex task (e.g., Tiny-ImageNet). The average difference of $Acc_{bucket}$ and $Acc_{confident}$ of Tiny-ImageNet is greater than 7\%, which is a very big bias. To understand why $Acc_{Aries}$ works, we check the results of $Acc_{bucket}$ and $Acc_{confident}$ separately and find that, generally, the $Acc_{bucket}$ is over-estimation (48 out of 52 cases) while the $Acc_{confident}$ is under-estimation (48 out of 52 cases). We conjecture that this is because the learned decision boundary can not thoroughly work well on the data that have shifted distribution. The potential reason is that, in fact, the performance of the model on the high confident data ($LVR$ is 1) of the shifted data can be lower than the original test data, e.g., 99.57\% (original test data) vs 95.10\% (Brightness data) of CIFAR-10-VGG16 model. The data with high confidence have a large proportion over all the data (e.g., there are 7969 data whose label consistent time is 50 for CIFAR-10-VGG16). Then, the results (line 4 in algorithm~\ref{alg:pe}) can be overestimated. On the other hand, there are more data that the model has low confidence in but are still correctly predicted, e.g., 43, 51, 44 (Brightness data) vs 24, 27, 18 (original test data) when $LVR$ times are 0.6, 0.62, and 0.64 of CIFAR-10-VGG16 model. This can make the size of high confident data in the shifted set as well as the $Acc_{confident}$ under-estimation. However, $Acc_{Aries}$ finally averages the under- and over-estimation and produces a more precise accuracy. In the remaining experiments, we only report the results of $Acc_{Aries}$. 

\begin{figure}[!h]
    \centering
    \subfigure[CIFAR10, ResNet20]{
    \includegraphics[scale=0.26]{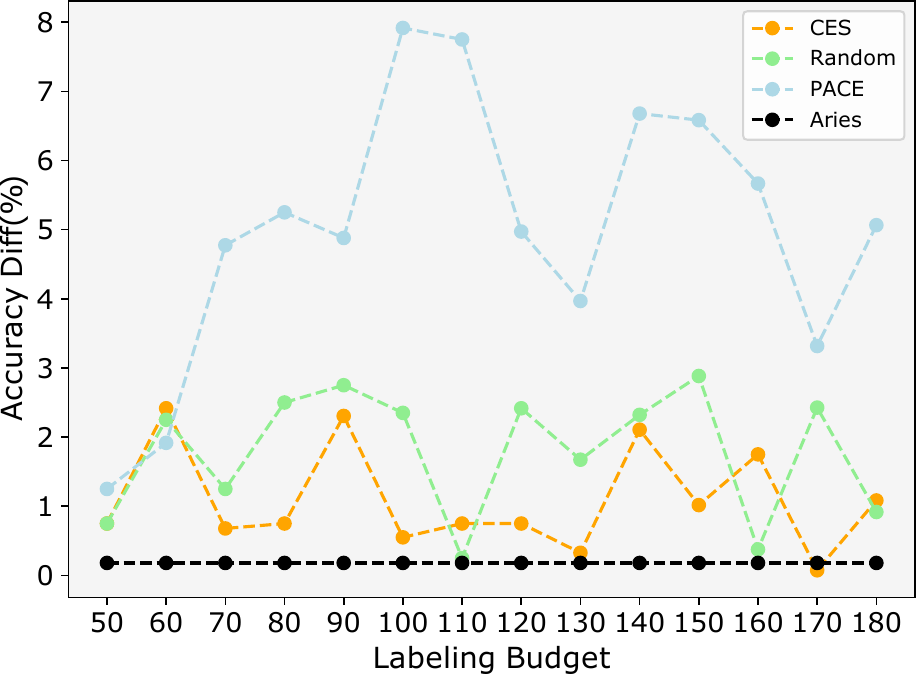}%
    }
    \subfigure[CIFAR10, VGG16]{
    \includegraphics[scale=0.26]{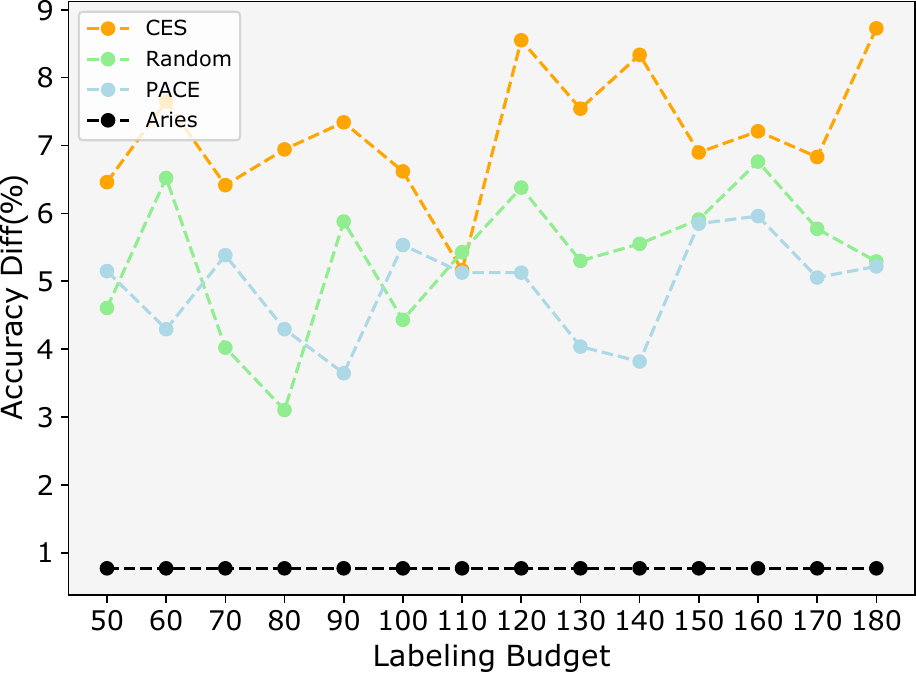}%
    }
    \subfigure[ImageNet, ResNet]{
    \includegraphics[scale=0.26]{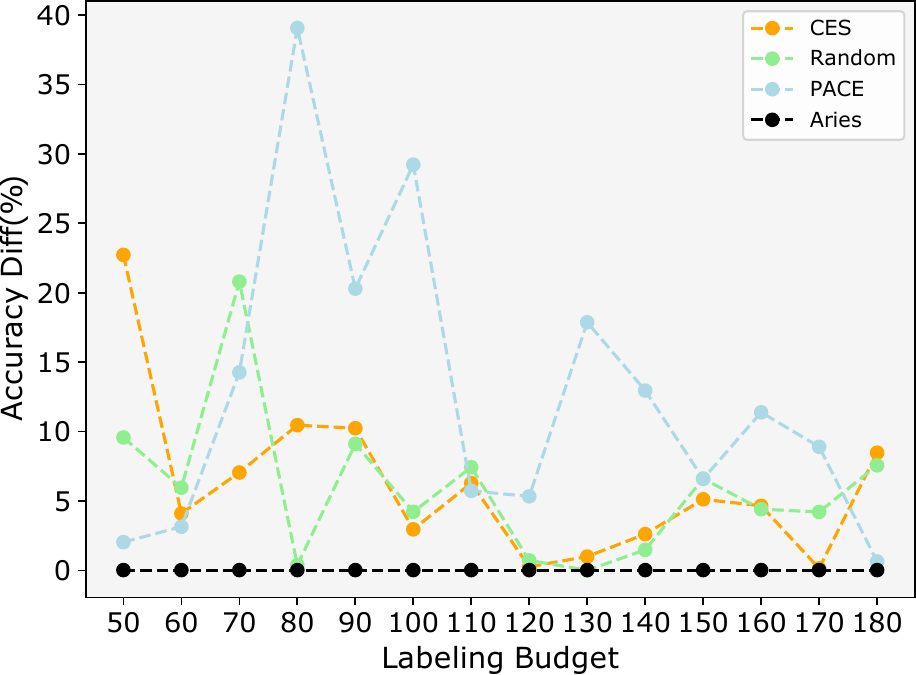}
    }
    \subfigure[ImageNet, DenseNet]{
    \includegraphics[scale=0.26]{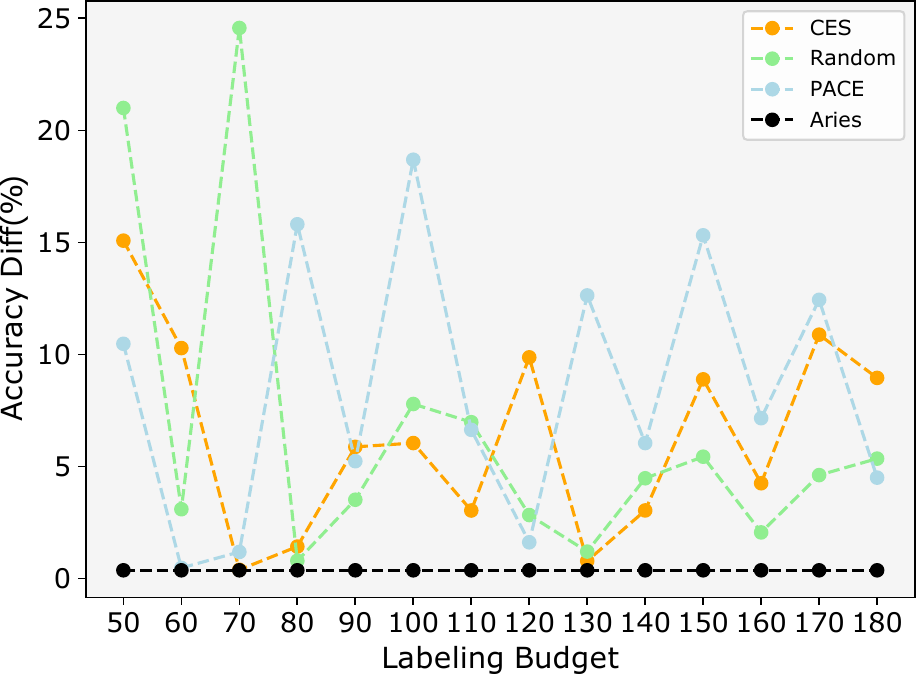}%
    }
    \caption{Comparison with test selection methods. Transformation: contrast.}
    \label{fig:rq1_selection}
\end{figure}

\textbf{Comparison with baselines.} First, we compare \textit{Aries} with labeling-free methods. From Table~\ref{table:rq1_robustness} we can see that \textit{Aries} outperforms the different configurations of the two baselines in most cases (50 out of 52). Besides, considering the average results, \textit{Aries} can always stand out.  Then, we compare \textit{Aries} with test selection methods. Fig.~\ref{fig:rq1_selection} presents the results of shifted test sets with the contrast transformation. Considering the results produced by each test selection method, we found that there are some conflicting conclusions compared to the original papers. For example,~\cite{chen2020practical} reports that PACE outperforms CES in its evaluation settings. However, from our results, only in CIFAR-10-VGG16-Contrast, PACE significantly outperforms CES. Under other settings, the results of these two methods fluctuate greatly. The same conflict occurs in random selection. In our evaluation, the existing well-designed test selection methods cannot consistently perform better than the random selection. This phenomenon reflects that, the evaluation of existing test selection methods for accuracy estimation is insufficient. Distribution shifts in data should be considered.

By comparison, \textit{Aries} achieves competitive results with test selection methods. Although in some situations, selection-based methods achieve better results than \textit{Aries}, e.g., in CIFAR-10-ResNet20-Brightness, when the labeling budget is 90, CES can estimate the accuracy more precisely. Overall, our technique performs the best in most cases (96 out of 128). Besides, \textit{Aries} achieves more stable performance than the selection-based methods. For example, in CIFAR-10-ResNet20-Brightness, the estimation bias of CES can vary from 0.01\% to 2.44\% by using different labeling budgets, which could waste human resources while obtaining unexpected results. 

\noindent\\\colorbox{gray!20}{\framebox{\parbox{0.96\linewidth}{
\textbf{Answer to RQ1}: \textit{Aries} estimates the model accuracy with a slight bias ranging from 0.03\% to 2.60\%. In addition, \textit{Aries} outperforms labeling-free methods in 50 out of 52 cases and test selection-based methods in 96 out of 128 cases. }}}

\subsection{RQ2: Influencing Factor Study}
\label{sec:rq2}

Next, we explore how different configurations and settings affect the performance of \textit{Aries}. As mentioned in Section~\ref{sec:pe}, there are three main factors we need to consider, the number of buckets ($n$ in algorithm~\ref{alg:pe}), the dropout rate of the dropout layer, and the distance approximation method. 

\begin{figure}[!h]
    \centering
    \subfigure[CIFAR10-ResNet20]{
    \includegraphics[scale=0.25]{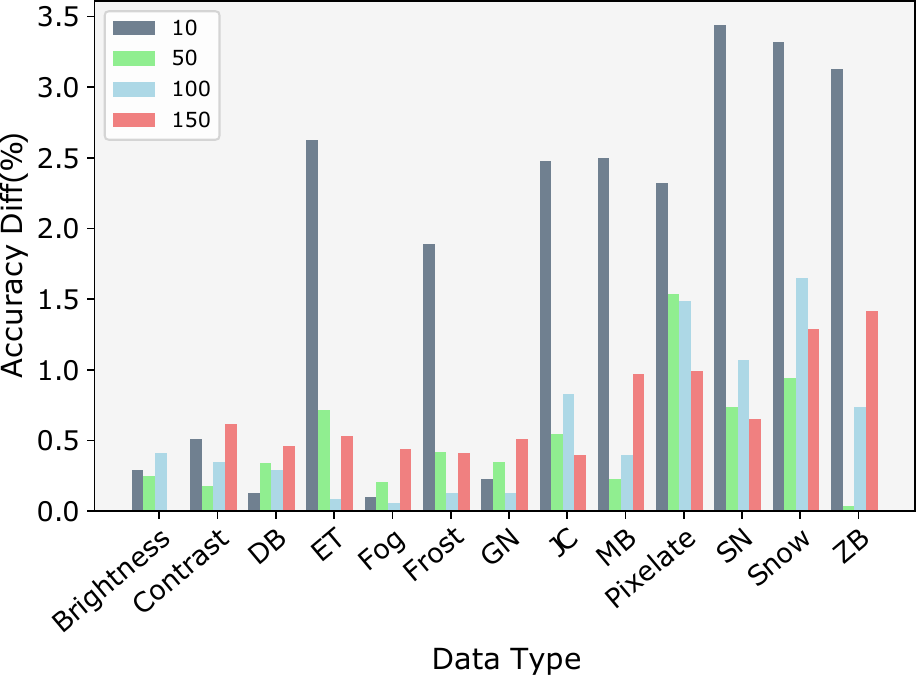}%
    }
    \subfigure[CIFAR10-VGG16]{
    \includegraphics[scale=0.25]{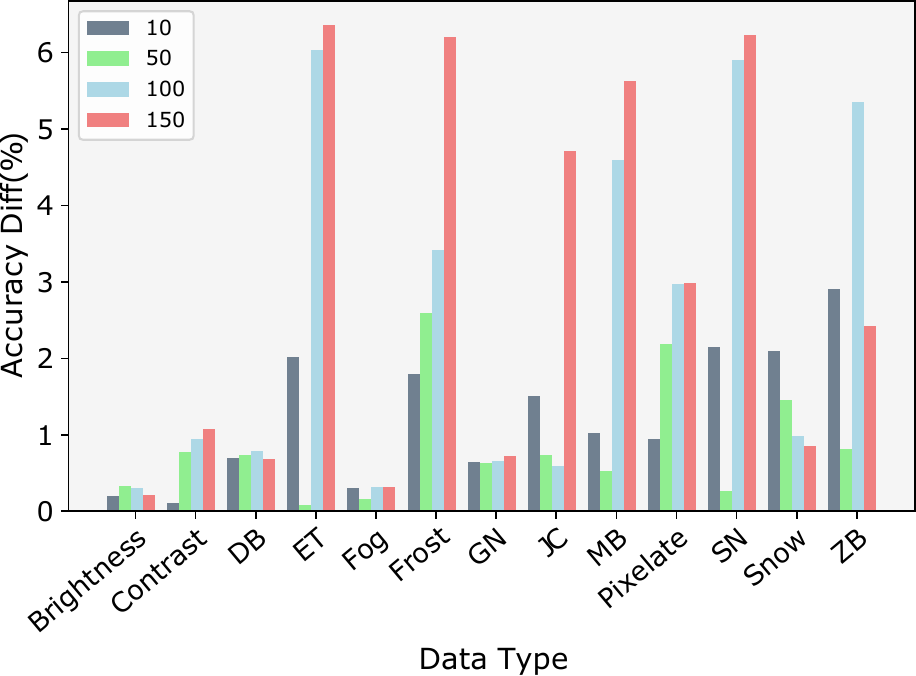}%
    }
    \subfigure[ImageNet-ResNet101]{
    \includegraphics[scale=0.25]{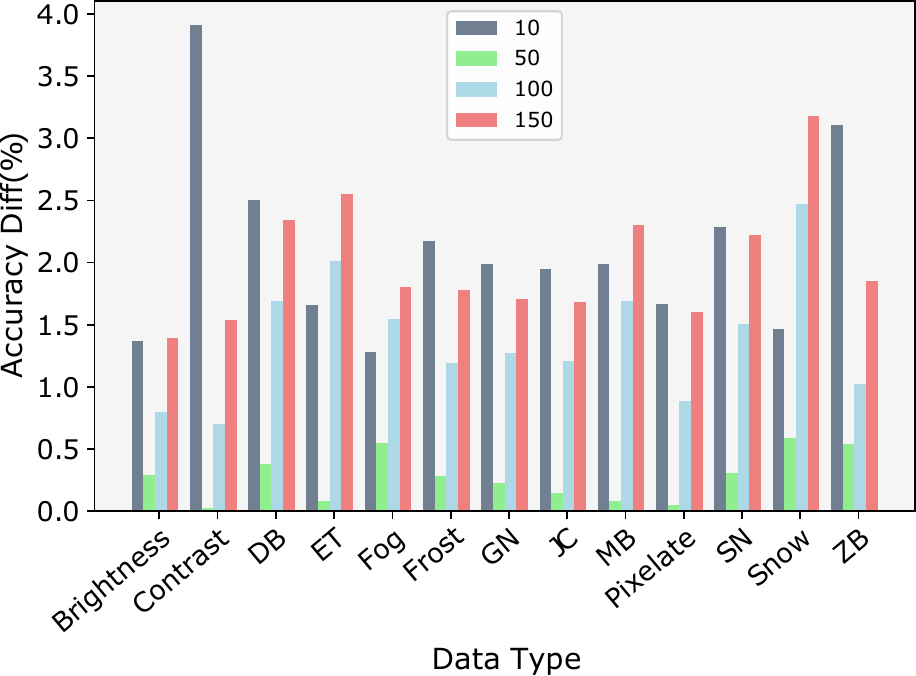}%
    }
    \subfigure[ImageNet-DenseNet121]{
    \includegraphics[scale=0.25]{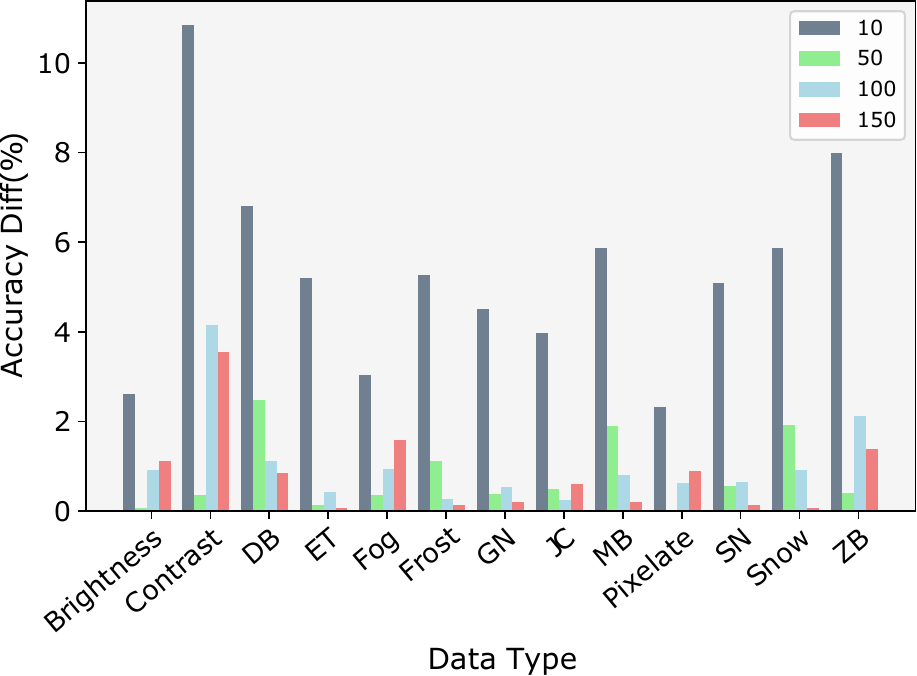}%
    }
    \caption{Accuracy difference (\%) between the real accuracy and the estimated accuracy by using different $Bucket$ numbers.}
    \label{fig:rq2_sec_num}
\end{figure}

\textbf{Number of Buckets.} Fig.~\ref{fig:rq2_sec_num} presents the results of the accuracy estimation using \textit{Aries} by different settings of the bucket number. The first conclusion we can draw is that, this factor has quite an impact on the results. For instance, in ImageNet-DenseNet121-Contrast, using 10 buckets can increase the accuracy difference by almost 10\% than using 50 buckets. However, it's clear that there is no such setting that performs consistently better than others in all datasets and models. For example, in CIFAR10-VGG16, $n=10$ is a relatively good setting. However, in CIFAR10-ResNet20, $n=10$ is the worst among the four settings which means the setting of $Bucket$ number is highly datasets and model-dependent. But if we check the more detailed values, we can see that, in total, in 5, 28, 6, and 13 cases, using 10, 50, 100, and 150 buckets can achieve the best estimation results, respectively. And on average, the differences between the estimated accuracy and the real accuracy are 2.62\%, 0.61\%, 1.39\%, and 1.61\%, respectively. Therefore, for the number of buckets, even though there can be no best setting, 50 is recommended among the studied settings. \textbf{Conclusion:} The number of $Bucket$ highly impacts the performance of \textit{Aries}. However, this hyperparameter setting is dataset and model-dependent. Thus, users should set this number according to the real use cases. $n=50$ is a default setting of \textit{Aries}.

\begin{table*}[!ht]
\centering
\caption{Difference (\%) between the real and estimated accuracy by \textit{Aries} using different dropout rates. DR: Dropout rate.}
\label{tab:drop_rate}
\resizebox{\textwidth}{!}
{
\begin{tabular}{lllccccccccccccc|c}
\toprule
\textbf{Dataset} & \textbf{DNN} & \textbf{DR} & \textbf{Brightness} & \textbf{Contrast} & \textbf{DB} & \textbf{ET} & \textbf{Fog} & \textbf{Frost} & \textbf{GN} & \textbf{JC} & \textbf{MB} & \textbf{Pixelate} & \textbf{SN} & \textbf{Snow} & \textbf{ZB} & \multicolumn{1}{c}{\textbf{Avg.}} \\ \hline
\multicolumn{1}{c}{} &  & \textbf{0.1} & 0.31 & 1.27 & 0.10 & 4.79 & 0.44 & 3.33 & \cellcolor[HTML]{C0C0C0}0.08 & 3.91 & 5.82 & 3.20 & 5.60 & 4.84 & 6.40 & 3.08 \\
\multicolumn{1}{c}{} &  & \textbf{0.2} & 0.26 & 0.94 & \cellcolor[HTML]{C0C0C0}0.03 & 3.53 & 0.39 & 2.64 & 0.11 & 3.12 & 3.99 & 2.66 & 4.43 & 4.03 & 4.67 & 2.37 \\
\multicolumn{1}{c}{} &  & \textbf{0.3} & \cellcolor[HTML]{C0C0C0}0.15 & 0.45 & 0.37 & 2.45 & 0.20 & 1.88 & 0.33 & 2.39 & 2.58 & 2.29 & 3.40 & 3.21 & 3.04 & 1.75 \\
\multicolumn{1}{c}{} &  & \textbf{0.4} & 0.19 & \cellcolor[HTML]{C0C0C0}0.14 & 0.29 & 1.55 & \cellcolor[HTML]{C0C0C0}0.07 & 1.16 & 0.30 & 1.75 & 1.28 & 1.86 & 2.46 & 2.47 & 1.67 & 1.17 \\
\multicolumn{1}{c}{} &  & \textbf{0.5} & 0.25 & 0.18 & 0.34 & 0.72 & 0.21 & 0.42 & 0.35 & 0.55 & \cellcolor[HTML]{C0C0C0}0.23 & 1.54 & 0.74 & 0.94 & \cellcolor[HTML]{C0C0C0}0.04 & \cellcolor[HTML]{C0C0C0}0.50 \\
\multicolumn{1}{c}{} &  & \textbf{0.6} & 0.27 & 0.58 & 0.37 & \cellcolor[HTML]{C0C0C0}0.51 & 0.35 & \cellcolor[HTML]{C0C0C0}0.36 & 0.27 & 0.42 & 1.37 & 1.33 & \cellcolor[HTML]{C0C0C0}0.52 & 1.42 & 1.48 & 0.71 \\
\multicolumn{1}{c}{} &  & \textbf{0.7} & 0.38 & 1.22 & 0.55 & 2.38 & 0.90 & 1.50 & 0.48 & \cellcolor[HTML]{C0C0C0}0.34 & 3.14 & 0.92 & 0.61 & 0.80 & 3.67 & 1.30 \\
\multicolumn{1}{c}{} &  & \textbf{0.8} & 0.26 & 2.68 & 0.71 & 5.13 & 1.56 & 3.33 & 0.56 & 2.22 & 5.91 & \cellcolor[HTML]{C0C0C0}0.16 & 2.36 & \cellcolor[HTML]{C0C0C0}0.73 & 7.00 & 2.51 \\
\multicolumn{1}{c}{} & \multirow{-9}{*}{\textbf{ResNet20}} & \textbf{0.9} & 0.28 & 5.70 & 1.63 & 10.76 & 3.91 & 8.06 & 1.57 & 6.40 & 11.21 & 1.65 & 6.66 & 5.30 & 14.16 & 5.95 \\ \cline{2-17} 
\multicolumn{1}{c}{} &  & \textbf{0.1} & 0.09 & \cellcolor[HTML]{C0C0C0}0.03 & 0.24 & \cellcolor[HTML]{FFFFFF}0.15 & 0.11 & \cellcolor[HTML]{C0C0C0}0.56 & \cellcolor[HTML]{C0C0C0}0.09 & 0.99 & \cellcolor[HTML]{C0C0C0}0.23 & 0.55 & 0.82 & 0.79 & \cellcolor[HTML]{C0C0C0}0.55 & \cellcolor[HTML]{C0C0C0}0.40 \\
\multicolumn{1}{c}{} &  & \textbf{0.2} & 0.25 & 0.23 & \cellcolor[HTML]{C0C0C0}0.13 & 1.37 & 0.03 & 0.89 & 0.25 & \cellcolor[HTML]{C0C0C0}0.56 & 0.70 & \cellcolor[HTML]{C0C0C0}0.31 & 0.73 & 1.03 & 1.85 & 0.64 \\
\multicolumn{1}{c}{} &  & \textbf{0.3} & 0.08 & 0.09 & 0.50 & 2.29 & \cellcolor[HTML]{C0C0C0}0.02 & 1.78 & 0.61 & 1.47 & 1.09 & 0.83 & 2.09 & 2.03 & 2.97 & 1.22 \\
\multicolumn{1}{c}{} &  & \textbf{0.4} & \cellcolor[HTML]{C0C0C0}0.04 & 0.29 & 0.58 & 3.28 & 0.13 & 3.12 & 0.46 & 2.41 & 2.02 & 1.41 & 0.52 & \cellcolor[HTML]{C0C0C0}0.00 & 4.01 & 1.41 \\
\multicolumn{1}{c}{} &  & \textbf{0.5} & 0.33 & 0.77 & 0.74 & \cellcolor[HTML]{C0C0C0}0.08 & 0.16 & 2.60 & 0.63 & 0.73 & 0.53 & 2.19 & \cellcolor[HTML]{C0C0C0}0.27 & 1.45 & 0.81 & \cellcolor[HTML]{FFFFFF}0.87 \\
\multicolumn{1}{c}{} &  & \textbf{0.6} & 0.07 & 0.37 & 1.44 & 6.66 & 0.13 & 6.15 & 0.83 & 4.96 & 5.28 & 3.46 & 6.20 & 8.14 & 8.01 & 3.98 \\
\multicolumn{1}{c}{} &  & \textbf{0.7} & 1.37 & 1.19 & 0.13 & 11.16 & 0.93 & 9.28 & 1.00 & 7.87 & 9.80 & 4.50 & 7.65 & 11.22 & 12.75 & 6.07 \\
\multicolumn{1}{c}{} &  & \textbf{0.8} & 2.14 & 7.92 & 3.09 & 17.70 & 6.21 & 14.39 & 1.31 & 14.03 & 16.51 & 6.66 & 10.76 & 15.33 & 22.79 & 10.68 \\
\multicolumn{1}{l}{\multirow{-18}{*}{\textbf{CIFAR-10}}} & \multirow{-9}{*}{\textbf{VGG16}} & \textbf{0.9} & 4.44 & 18.12 & 20.54 & 27.27 & 1.19 & 16.61 & 7.01 & 20.87 & 25.63 & 0.79 & 10.75 & 24.61 & 30.86 & 16.05 \\ \hline
 &  & \textbf{0.1} & 3.12 & 7.44 & 5.14 & 4.86 & 3.74 & 4.46 & 5.02 & 4.09 & 4.90 & 3.85 & 4.97 & 4.02 & 6.18 & 4.75 \\
 &  & \textbf{0.2} & 2.36 & 5.24 & 3.33 & 3.01 & 2.25 & 3.01 & 3.60 & 2.72 & 2.78 & 2.76 & 3.42 & 2.31 & 4.19 & 3.15 \\
 &  & \textbf{0.3} & 1.37 & 3.23 & 1.73 & 1.36 & 0.75 & 1.81 & 2.02 & 1.23 & 1.42 & 1.57 & 1.78 & 0.82 & 2.57 & 1.67 \\
 &  & \textbf{0.4} & 0.18 & 1.29 & \cellcolor[HTML]{C0C0C0}0.11 & 0.60 & \cellcolor[HTML]{C0C0C0}0.22 & 0.37 & 0.29 & 0.24 & \cellcolor[HTML]{C0C0C0}0.06 & 0.36 & \cellcolor[HTML]{C0C0C0}0.06 & \cellcolor[HTML]{C0C0C0}0.53 & 0.56 & 0.37 \\
 &  & \textbf{0.5} & 0.29 & \cellcolor[HTML]{C0C0C0}0.03 & 0.38 & \cellcolor[HTML]{C0C0C0}0.08 & 0.55 & \cellcolor[HTML]{C0C0C0}0.28 & \cellcolor[HTML]{C0C0C0}0.23 & \cellcolor[HTML]{C0C0C0}0.15 & 0.08 & \cellcolor[HTML]{C0C0C0}0.05 & 0.31 & 0.59 & \cellcolor[HTML]{C0C0C0}0.54 & \cellcolor[HTML]{C0C0C0}0.27 \\
 &  & \textbf{0.6} & 1.11 & 1.44 & 1.69 & 2.70 & 2.01 & 1.65 & 1.44 & 1.75 & 2.14 & 1.30 & 2.36 & 3.08 & 1.68 & 1.87 \\
 &  & \textbf{0.7} & 1.82 & 3.10 & 2.92 & 3.36 & 2.82 & 2.66 & 2.94 & 3.15 & 2.72 & 2.56 & 3.80 & 4.26 & 3.37 & 3.04 \\
 &  & \textbf{0.8} & \cellcolor[HTML]{C0C0C0}0.16 & 3.95 & 3.30 & 4.03 & 2.32 & 4.16 & 3.03 & 2.52 & 2.83 & 1.86 & 4.04 & 3.99 & 3.36 & 3.04 \\
 & \multirow{-9}{*}{\textbf{ResNet101}} & \textbf{0.9} & 5.81 & 2.88 & 3.60 & 8.32 & 4.12 & 1.02 & 1.09 & 1.13 & 1.86 & 3.06 & 3.37 & 6.34 & 0.83 & 3.34 \\ \cline{2-17} 
 &  & \textbf{0.1} & 0.24 & 5.41 & 2.15 & 1.40 & 0.28 & 1.03 & 1.04 & \cellcolor[HTML]{FFFFFF}0.30 & 1.45 & 0.24 & 1.26 & 1.76 & 3.20 & 1.52 \\
 &  & \textbf{0.2} & 0.09 & \cellcolor[HTML]{C0C0C0}0.34 & 2.56 & \cellcolor[HTML]{FFFFFF}0.05 & 0.55 & 1.11 & 1.66 & 0.38 & 1.49 & 0.17 & 1.40 & 1.78 & 1.36 & 1.00 \\
 &  & \textbf{0.3} & 0.09 & 1.52 & 2.41 & 0.31 & 0.52 & \cellcolor[HTML]{C0C0C0}0.84 & 1.36 & 0.34 & 1.62 & 0.14 & 1.62 & 2.08 & 1.09 & 1.07 \\
 &  & \textbf{0.4} & 0.36 & 0.83 & 2.71 & 0.46 & \cellcolor[HTML]{C0C0C0}0.02 & 1.45 & 1.60 & 0.75 & 1.84 & 0.15 & 1.76 & 2.42 & 0.66 & 1.15 \\
 &  & \textbf{0.5} & \cellcolor[HTML]{C0C0C0}0.07 & 0.37 & 2.49 & 0.15 & 0.36 & 1.13 & \cellcolor[HTML]{C0C0C0}0.38 & 0.50 & 1.90 & \cellcolor[HTML]{C0C0C0}0.03 & \cellcolor[HTML]{C0C0C0}0.56 & 1.92 & 0.41 & \cellcolor[HTML]{C0C0C0}0.79 \\
 &  & \textbf{0.6} & 0.37 & 0.78 & 2.34 & 0.42 & 0.38 & 1.30 & 1.77 & 0.63 & 1.85 & 0.06 & 1.85 & 1.65 & \cellcolor[HTML]{C0C0C0}0.33 & 1.06 \\
 &  & \textbf{0.7} & 0.21 & 0.55 & 2.39 & \cellcolor[HTML]{C0C0C0}0.01 & 0.60 & 1.00 & 1.08 & 0.35 & 1.57 & 0.35 & 2.01 & 1.92 & 1.07 & 1.01 \\
 &  & \textbf{0.8} & 0.46 & 0.39 & \cellcolor[HTML]{C0C0C0}2.15 & 0.15 & 0.37 & 0.98 & 1.35 & 0.33 & \cellcolor[HTML]{C0C0C0}1.43 & 0.62 & 1.33 & \cellcolor[HTML]{C0C0C0}1.64 & 1.45 & 0.97 \\
\multirow{-18}{*}{\textbf{Tiny-ImageNet}} & \multirow{-9}{*}{\textbf{DenseNet121}} & \textbf{0.9} & 0.08 & 0.65 & 2.51 & 0.29 & 0.60 & 0.97 & 1.16 & \cellcolor[HTML]{C0C0C0}0.23 & 1.58 & 0.01 & 1.95 & 1.94 & 1.52 & 1.04 \\ \toprule
\end{tabular}
}
\end{table*}

\begin{figure}[!h]
	\centering
	\includegraphics[width=0.4\textwidth]{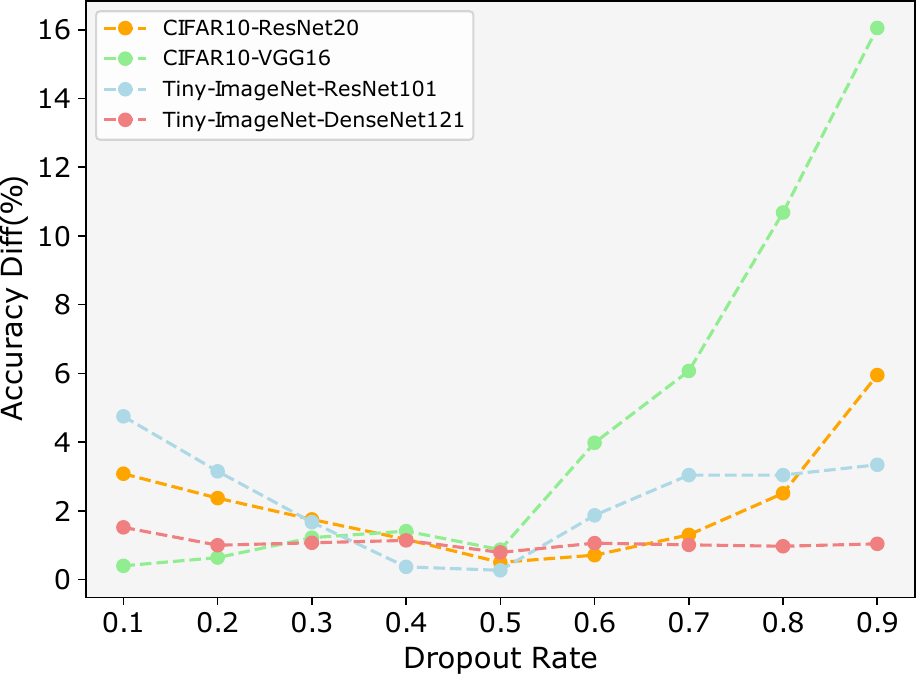}
	\caption{The trend of the average difference between real accuracy and estimated accuracy by using different dropout rates. The rate at 0.5 is a common turning point of the accuracy difference for all datasets and models.}
	\label{fig:drop_tresnding}
\end{figure}

\textbf{Dropout Rate.} Table~\ref{tab:drop_rate} presents the difference between the real and estimated accuracy by \textit{Aries} using different dropout rates. Similar to the study of the $Bucket$ number, there is no dropout rate setting that can consistently outperform others. But still, a relatively better setting exists through a deeper analysis. In total, in 6, 5, 3, 10, 19, 4, 2, 6, and 1 cases, the dropout rate 0.1, 0.2, 0.3, 0.4, 0.5, 0.6, 0.7, 0.8, and 0.9 achieve the best estimation results, respectively. This indicates that the dropout rate = 0.5 is the best among all the studied settings in terms of achieving the most precise estimation. Then, looking into the average results, we draw a similar conclusion that when we set the dropout rate as 0.5, in 3 (out of 4) cases, the estimated accuracy is closer to real accuracy compared to other settings. Fig.~\ref{fig:drop_tresnding} depicts the trend of the average difference between real and estimated accuracy by using different dropout rates (Column \textit{Average} in Table~\ref{tab:drop_rate}). We can see that, the difference drops first when the dropout rate increases, and after the dropout rate reaches around 0.5, the difference increases. \textbf{Conclusion: } There is no dropout rate setting that always performs the best. We recommend using the dropout rate of around 0.5 for \textit{Aries} to gain better results.

\begin{table}[!ht]
\caption{Comparison between dropout and mutant. Each value is the absolute difference between real and estimated accuracy (\%).}
\label{tab:mutation_results}
\resizebox{\columnwidth}{!}
{
\begin{tabular}{lcccc|cccc}
\hline
 & \multicolumn{4}{c|}{\textbf{CIFAR-10}} & \multicolumn{4}{c}{\textbf{Tiny-ImageNet}} \\ \cline{2-9}
 & \multicolumn{2}{c}{\textbf{ResNet20}} & \multicolumn{2}{c|}{\textbf{VGG16}} & \multicolumn{2}{c}{\textbf{ResNet101}} & \multicolumn{2}{c}{\textbf{DesNet121}} \\
 \multirow{-3}{*}{\textbf{Data Type}}& \textbf{Mutant} & \textbf{Dropout} & \textbf{Mutant} & \textbf{Dropout} & \textbf{Mutant} & \textbf{Dropout} & \textbf{Mutant} & \textbf{Dropout} \\ \hline
\textbf{Brightness} & 0.12 & 0.25 & 0.01 & 0.33 & 4.49 & 0.29 & 6.71 & 0.07 \\
\textbf{Contrast} & 0.95 & 0.18 & 0.00 & 0.77 & 10.67 & 0.03 & 19.62 & 0.37 \\
\textbf{DB} & 0.24 & 0.34 & 0.15 & 0.74 & 7.80 & 0.38 & 14.08 & 2.49 \\
\textbf{ET} & 3.30 & 0.72 & 1.19 & 0.08 & 7.72 & 0.08 & 11.74 & 0.15 \\
\textbf{Fog} & 0.51 & 0.21 & 0.14 & 0.16 & 6.02 & 0.55 & 8.43 & 0.36 \\
\textbf{Frost} & 2.97 & 0.42 & 2.36 & 2.60 & 7.08 & 0.28 & 11.25 & 1.13 \\
\textbf{GN} & 0.25 & 0.35 & 0.21 & 0.63 & 7.28 & 0.23 & 9.94 & 0.38 \\
\textbf{JC} & 3.38 & 0.55 & 2.68 & 0.73 & 6.39 & 0.15 & 9.52 & 0.50 \\
\textbf{MB} & 4.04 & 0.23 & 1.17 & 0.53 & 7.34 & 0.08 & 12.77 & 1.90 \\
\textbf{Pixelate} & 2.80 & 1.54 & 1.56 & 2.19 & 5.43 & 0.05 & 6.43 & 0.03 \\
\textbf{SN} & 4.49 & 0.74 & 3.00 & 0.27 & 7.34 & 0.31 & 10.67 & 0.56 \\
\textbf{Snow} & 3.66 & 0.94 & 2.95 & 1.45 & 6.67 & 0.59 & 11.61 & 1.92 \\
\textbf{ZB} & 4.50 & 0.04 & 1.26 & 0.81 & 9.28 & 0.54 & 15.50 & 0.41 \\ \hline
\textbf{Avg.} & 2.40 & 0.50 & 1.28 & 0.87 & 7.19 & 0.27 & 11.41 & 0.79 \\ \hline
\end{tabular}
}
\end{table}

\textbf{Mutant for Distance Estimation.} Finally, since at each prediction time, the dropout model can be seen as a mutant of the original model, we explore if we can utilize model mutation for replacing dropout prediction to approximate the distance between data and decision boundaries.

Table~\ref{tab:mutation_results} presents the results of the comparison between the two ways of approximating the distance between the data and the decision boundaries. When replacing the dropout with mutants (change $M$ in Definition~\ref{def: lvr} to mutants), \textit{Aries} still works in some cases. For CIFAR-10, \textit{Aries} with mutants can still produce some acceptable results, e.g., in 10 cases, the difference is smaller than 1\%. Compared to \textit{Aries} with dropout, in 10 out of 26 cases, the mutant achieves better results. On average, there are only 1.9\% and 0.42\% effectiveness gaps between these two ways. However, the performance of \textit{Aries} with mutants becomes worse in Tiny-ImageNet, and the difference increases significantly compared to using dropout. In all the cases, \textit{Aries} with mutants performs worse than with dropout. This phenomenon indicates that \textit{Aries} with mutants can only work on simple datasets and models. This is reasonable because, at each prediction time, the status of the dropout model might be appeared in the training process due to its design nature~\cite{srivastava2014dropout}. Thus, it can reflect the learned decision boundary more precisely. However, the post-training model mutation randomly modifies the model, which could totally change the learned decision boundary even if it maintains the accuracy. \textbf{Conclusion: } \textit{Aries} with using mutants can achieve similar accuracy estimation results with using dropout in CIFAR-10 dataset, while fails in Tiny-ImageNet dataset. It needs more careful mutation operator selection and hyperparameter tuning to ensure the decision boundary does not change too much after model mutation.

\noindent\\\colorbox{gray!20}{\framebox{\parbox{0.96\linewidth}{\textbf{Answer to RQ2}: The number of buckets and dropout rate affect the performance of \textit{Aries}. 50 and 0.5 are the recommended settings, respectively. Simply replacing the dropout with mutant can work on the simple dataset (CIFAR-10) but fails on the complex dataset (Tiny-ImageNet). }}}

\section{Discussion and Threat to Validity}
\label{sec:discussion}

\subsection{Limitations \& Future Directions}
\textbf{Limitations.} 1) The first potential limitation is that \textit{Aries} might suffer from adversarial attacks~\cite{ren2020adversarial}. A strong adversarial attack method can control the distance between the adversarial examples and the decision boundaries by pushing the original data close to or far away from the decision boundary, which invalidates our learned boundary information. However, adversarial attack is a common concern for all 
methods. For example, for the output-based methods (PACE and CES), white-box adversarial attacks can be designed to change the output of the neurons to force these methods to select useless data to label. How to defend against adversarial attacks is an open problem. 2) The second limitation comes from our assumption that we already have some labeled test data. In general, this assumption can stand. However, in extreme cases where only the model and new unlabeled data are available, we still need to undertake data labeling. 

\textbf{Future directions.} 1) We only utilize the original labeled test data to gain the decision boundary information, and it works well in our evaluation subjects. There could be a way to do data augmentation based on the labeled data and increase the data space we have. In this way, we can learn more precise boundary information and, therefore, make better accuracy estimations for the new unlabeled data. 2) Although dropout uncertainty is the widely studied uncertainty method and works well in our technique, the mutant prediction is the closest way to the dropout prediction. Some other uncertainty methods can be used to approximate the distance between data to the decision boundaries, e.g., DeepGini~\cite{feng2020deepgini}. We plan to study more methods and explore if there is a better way to replace dropout prediction. On the other hand, adversarial attacks can be used to achieve the same goal~\cite{malanie2018dfal}. 3) Maybe more interestingly, we tend to explore if \textit{Aries} can be used in other types of datasets and models, e.g., models for code learning.

\subsection{Threats to Validity}
The internal threats to validity are the implementations of our technique, the baselines, and mutation operators as well as the preparation of new unlabeled data. Our technique is simple and easy to implement and its core part is using pure Python. Also, we release our code for future study. All the implementations of the baselines and the mutation operators are from the original papers. For the new unlabeled data, to reduce the influence of parameter settings (e.g., which levels of brightness should be added), we directly reuse the released datasets that are widely studied in the literature. 

The external threats to validity come from the selected datasets and models for our evaluation. For the dataset, we use two commonly studied ones from the recent research~\cite{cui2021learnable, bengar2021reducing, leino2021globally}. For each dataset, we build two different model architectures from simple to complex. Compared to the existing test selection works~\cite{chen2020practical, li2019boosting} which stop by ResNet-50, our considered model architectures are more complex (ResNet101 and DenseNet121). Besides, another threat that comes from the model could be model calibration~(roughly speaking, the diversity of models)~\cite{nixon2019measuring}. Actually, our evaluation involves both well-calibrated and poorly-calibrated models. Indeed, the Predicted score-based method can witness model calibration to some extent. For CIFAR-10-VGG16, this method (with $\tau$ = 0.9) reveals that around 90\% of data have \textgreater\ 90\% confidence, thus, the model is over-confidence. For a similar reason, Tiny-ImageNet-ResNet101 is under-confident. And for the new unlabeled data, we also follow the previous works~\cite{hendrycks2019augmix, hendrycks2021pixmix, kang2019testing} that evaluate the model robustness to prepare our test sets. Even though collecting unlabeled data in the wild is a good way to further evaluate our method. It is not easy to collect and label massive new data. In this paper, we believe the 13 transformation techniques we used to simulate distribution shifts can achieve high data diversity.  

\section{Related Work}
\label{sec:related_work}

\subsection{Coverage Design and Automated Testing of DNNs}
Recently, DNN testing has become a very active research area~\cite{braiek2019deepevolution, birchler2021automated, panichella2021we, zhou2020cost}, with quite a few techniques proposed to ensure the quality of DNN from different angles. As a very early testing technique, DeepXplore, proposed by Pei \emph{et al.}, first defined the concept of neuron coverage and utilized it as a measure to generate test data to test DNN models. After that, Ma \emph{et al.} proposed DeepGauge, which contains more multi-granularity neuron coverage criteria, e.g., k-multisection Neuron Coverage. Then, more practically, DeepTest~\cite{tian2018deeptest} and TACTIC~\cite{li2021testing} utilize the search-based method by using neuron coverage as a search objective to synthesize test data to test the DNN-based autonomous driving systems. Fuzz testing~\cite{godefroid2008automated}, which is a famous testing technique in conventional software engineering, has also been applied for DL testing recently~\cite{guo2018dlfuzz}. Xie \emph{et al.} proposed DeepHunter~\cite{xie2019deephunter}, which generates test data by fuzzing while maximizing the neuron coverage. Different from other methods, DeepHunter contains seed selection and fuzzing as two phases. Besides, to enhance the model training and improve the robustness of trained models, Gao \emph{et al.} proposed SENSEI~\cite{gao2020fuzz} that can optimize the data augmentation process by the genetic algorithm at the model training time. Compared to these testing techniques that target proposing testing criteria and test data generation methods to help DL testing, our work focuses on efficiently testing DL models by performance estimation.  

\subsection{Test Selection for DNN}
As mentioned before, there are two categories of test selection methods, test selection for data prioritization and test selection for model performance estimation. For data prioritization, Kim \emph{et al.}~\cite{kim2019guiding} proposed two test adequacy methods, Likelihood-based Surprise Adequacy (LSA) and Distance-based Surprise Adequacy (DSA) by comparing the likelihood and distance between the training data and test data. They demonstrated these methods could be used to guide the retraining to produce more accurate models. Feng \emph{et al.}~\cite{feng2020deepgini} proposed DeepGini, a test prioritization technique based on the statistical perspective of the output of DNN models. The authors show that compared to the neuron coverage-based methods, DeepGini is a better technique for misclassified data indication and retraining guidance. More recently, Wang \emph{et al.}~\cite{prior2021} proposed a mutation-based test prioritization method that mutates both the input data and models to find the sensitive test data. On the other hand, for the test selection for model performance estimation, there are two works, PACE~\cite{chen2020practical}  and CES~\cite{li2019boosting}. We introduced these two methods in Section~\ref{sec:setup}. More recently, few works conducted empirical studies to explore the effectiveness of existing test selection methods in terms of fault detection~\cite{ma2021test} and performance of selection-based model repair~\cite{hu2022empirical}. Compared to the previous works, the main advantage of our technique is that \textit{Aries} does not need any extra data labeling effort, which makes the whole testing process for quality assessment of DNN automatic and less expensive.

\section{Conclusion}
\label{sec:conclusion}
In this paper, we proposed \textit{Aries} to automatically estimate the accuracy of DNN models on unlabeled data without labeling effort. The main intuition of \textit{Aries} is that a model should have similar prediction accuracy on the data that have similar distances to the decision boundaries. Specifically, \textit{Aries} employs the dropout uncertainty to approximate the distance between data and decision boundaries and learns this boundary information from the original labeled test data to estimate the accuracy of new unlabeled data. Our evaluation of two commonly studied datasets, four DNN architectures, and 13 types of unlabeled data demonstrated that \textit{Aries} can precisely predict the model accuracy. Besides, we demonstrated that \textit{Aries} outperforms SOTA labeling-free estimation methods and test selection-based methods. 

\section*{Acknowledgments}
This work is supported by the Luxembourg National Research Funds (FNR) through CORE project C18/IS/12669767/STELLAR/LeTraon.

\bibliographystyle{IEEEtran}
\balance
\bibliography{IEEEabrv,sample-base}

\end{document}